\documentclass{jfm}
\usepackage{graphicx}
\usepackage{epstopdf,epsfig}
\usepackage{amssymb}
\usepackage{amsmath}
\usepackage{subfig}
\usepackage{color,soul}
\usepackage{natbib}

\usepackage{framed}
\usepackage{xcolor}

 \def\vector#1{\mbox{\boldmath $#1$}}

\begin{document}

\newtheorem{lemma}{Lemma}
\newtheorem{corollary}{Corollary}

\shorttitle{Pairwise interactions of bubbles} 
\shortauthor{Maeda et al} 

\title{On viscid-inviscid interactions of a pair of bubbles rising near the wall}

\author
 {
 Kazuki Maeda\aff{1}
  \corresp{\email{kemaeda@stanford.edu}},
  Masanobu Date\aff{2},
  Kazuyasu Sugiyama\aff{3},\\
  Shu Takagi\aff{4}
  \and 
  Yoichiro Matsumoto\aff{4,5}
  }

\affiliation
{
\aff{1}
Center for Turbulence Research, Stanford University, Stanford, CA 94305, USA
\aff{2}
Central Japan Railway Company, 1-1-4 Meieki, Nakamura-ku, Nagoya, Aichi 450-6101, Japan
\aff{3}
Department of Mechanical Science and Bioengineering, Osaka University,\\ 1-3 Machikaneyama, Toyonaka, Osaka 560-0043, Japan
\aff{4}
Department of Mechanical Engineering, The University of Tokyo,\\ 7-3-1 Hongo, Bunkyo City, Tokyo 113-8656, Japan
\aff{5}
Tokyo University of Science, 1-3 Kagurazaka, Shinjuku City, Tokyo 162-8601, Japan
}
\maketitle

\begin{abstract}
Series of experiments on turbulent bubbly channel flows observed bubble clusters near the wall which can change large-scale flow structures. To gain insights into clustering mechanisms, we study the interaction of a pair of spherical bubbles rising in a vertical channel through combined experiments and modeling. Experimental imaging identifies that pairwise bubbles of 1.0 mm diameter take two preferred configurations depending on their mutual distance: side-by-side positions for a short distance ($S<5$) and nearly inline, oblique positions for a long distance ($S>5$), where $S$ is the mutual distance normalized by the bubble radius. In the model, we formulate the motions of pairwise bubbles rising at $Re=O(100)$. Analytical drag and lift, and semi-empirical, spatio-temporal stochastic forcing are employed to represent the mean acceleration and the fluctuation due to turbulent agitation, respectively. The model is validated against the experiment through comparing Lagrangian statistics of the bubbles. Simulations using this model identify two distinct timescales of interaction dynamics which elucidate the preferred configurations. For pairs initially in-line, the trailing bubble rapidly escapes from the viscous wake of the leading bubble to take the oblique position. Outside of the wake, the trailing bubble travels on a curve-line path with a slower velocity driven by potential interaction and horizontally approaches the leading bubble to become side-by-side. Moreover, statistical analysis identifies that the combination of the wake and the agitation can significantly accelerate the side-by-side clustering of in-line pairs. These results indicate positive contributions of liquid viscosity and turbulence to the formation of bubble clusters.
\end{abstract}


\section{Introduction}
Bubbly flows are of critical use in industrial applications including the enhancement and control of chemical processing, skin friction reduction, heat transfer in systems as diverse as chemical plants (\citet{Gong07,Hibiki02}), water vehicles and vessels \citep{Ceccio10, Gils13, Watamura13}, and nuclear reactors \citep{Tomiyama98}, as well as ubiquitous in nature as seen in breaking of oceanic waves (e.g. \citet{Thorpe80,Chan19,Chan20}) and aeration of waterfalls (e.g. \citet{Toombes00}).
Experiments on upward bubbly flow in vertical channels have reported that spherical bubbles tend to migrate away from the tunnel center and form a high-void fraction layer near the wall (\citet{Serizawa75, Wang87, Liu93}). This migration has been associated with the shear-induced lift force that acts on the bubbles to the direction toward which the velocity of the flow relative to the bubbles increases \citep{Auton87,Tomiyama02}. Such bubble clusters can interact with the wall boundary layer and alter the large-scale flow structures.
\citet{So02} reported laser Doppler velocimetry (LDV) measurements of upward bubbly flows in a vertical channel containing monodisperse, spherical bubbles rising at $Re=O(100)$ in water.
They reported that, when a surfactant is added to water, the coalescence of bubbles are prevented and that the bubbles accumulate near the channel wall to form crescent-like clusters perpendicular to the flow direction. Those clusters slide up faster than a single isolated bubble due to enhanced buoyancy, and lift up the surrounding fluid. The measurements also identified that, under the presence of the clusters, the mean flow velocity profile is steepened and the turbulent intensity is enhanced near the wall, leading to the reduction in the skin-friction drug. Using a similar experimental setup, \citet{Takagi07} identified a critical dependence of the cluster formation on the concentration and species of surfactants.
\citet{Fukuta08} numerically obtained the lift coefficient of a single bubble which is contaminated by surfactant adsorption causing the Marangoni effect.
Qualitatively, once accumulated near the wall, local hydrodynamic interactions among these bubbles have been considered to trigger the formation and growth of the bubble clusters.
Detailed mechanisms of the inter-bubble interactions that contribute to the onset of cluster formation remain largely elusive.

Inter-bubble interactions of spherical bubbles have long been studied based on potential theory.
\citet{Wijngaarden76} reported a model of equal-sized two spherical bubbles interacting in a perfect fluid. \citet{Biesheuvel82} used this model to numerically compute trajectories of a pair of bubbles in mutual interactions. \citet{Kok93} extended the model to include the effect of viscous diffusion by considering the global kinetic energy balance of \citet{Levich62}.
\citet{Harper70} predicted that pairwise bubbles rising in in-line configurations reach equilibrium positions where a balance is found between the attraction due to the drag reduction of the trailing bubble in the viscous wake of the leading bubble and the repulsion due to potential interaction. Both a direct numerical simulation by \citet{Yuan94} and an extended analysis by \citet{Harper97} reported the presence of the equilibrium distance for a pair of bubbles rising in-line at $0.2\leq Re\leq35$. \citet{Legendre03} modeled interactions of a pair of bubbles rising side-by-side in viscous liquid rising at moderate Reynolds numbers ($50\leq Re\leq500$), considering the viscous diffusion in the thin boundary layer at the surface of bubbles through direct numerical simulation (DNS) that resolves the layer.
\citet{Hallez11} extended this model for pairwise bubbles with arbitrary angle configurations to include the effect of the viscous wake of the leading bubble on the trailing bubble. This model predicts that the wake induces a shear-induced lift force on the trailing bubble in the direction outward from the wake.
\citet{Lu13} conducted DNS of an upward, turbulent bubbly flows containing $\sim$200 nearly spherical gas bubbles in a periodic channel rising at $Re=O(100)$, and quantified the modifications of the wall stress and the velocity profile of the flow due to the presence of bubble clusters.
\citet{Du20} conducted DNS of a similar number of spherical and deforming bubbles in a vertical channel, and observed clustering of spherical bubbles near the wall as well as quantified the inter-phase momentum transfer in the flow.

In experiments, \citet{Katz96} observed pairs of nearly spherical bubbles rising in-line at $0.2\leq Re\leq35$ in quiescent water, and reported that the in-line configurations are unstable and bubbles eventually collide with each other, against the aforementioned theoretical prediction. They associate this instability with deformations of bubbles.
\citet{Zenit01} observed weak clustering of bubbles rising at $Re=O(100)$ in a thin vertical channel, in a regime at which the bubble velocity is attenuated by frequent bubble-wall collisions.
\citet{Sanada05} observed a vertical bubble chain rising at $300\leq Re\leq600$ in quiescent liquid and identified that the bubbles tend to be scattered from the chain configuration.
\citet{Ogasawara18} observed the formations of clusters of monodisperse bubbles rising near the wall at $Re=O(100)$ in an inclined channel.
\citet{Kusuno19} used a setup similar to \citet{Sanada05} and reported that the inline configuration is unstable for pairwise bubbles rising at $10\leq Re\leq100$, and associated this instability with the wake-induced lift.

The purpose of the present study is to analyze the pairwise interaction of bubbles rising in turbulent channel bubbly flow through combined experiment and modeling, and gain insights into the onset mechanism of the cluster formation.
The dynamics of wall-bounded turbulent bubbly flows is multi-scale and rich in phenomenology. Measuring flow structures near the wall in the presence of the bubbly-layer is a challenging task. Direct computation of the entire spatial and temporal scales of the flow field associated with the bubble clustering can be prohibitively expensive even for the state-of-the-art high-performance computers, due to the requirement for simultaneously resolving the small-scale structures including the thin boundary layer at the surface of bubbles and capturing large-scale flow structures including $O(10^3)$ or more bubbles during a statistically significant timescale. Instead of taking such direct, inclusive approaches, we combine high-speed imaging and Lagrangian point-bubble modeling.
This approach effectively avoids the aforementioned difficulties.
The bubbles can be directly observed and tracked in high-speed images in a simple manner. The computational expense for this model is orders of magnitude smaller than the direct approach, enabling a large number of simulations for a sufficiently long period of time. In the model, we analytically express drag and lift coefficients of pairwise bubbles with arbitrary configurations. We derive the inviscid (potential) contribution to the drag and lift through bi-polar expansion of the velocity potential around the bubbles, and adopt the model of \citet{Hallez11} to express the contribution of the viscous effects. We model the fluctuation of the motions of bubbles due to turbulent agitation through empirical, spatio-temporal stochastic forcing of the bubbles. The spatio-temporal of this forcing are chosen such that the model reproduces Lagangian statistics of the bubbles obtained in the measurement a posteriori.
In the analysis, we focus on the relative trajectories of the pairwise bubbles as well as on the timescale of their motions, and quantify the effect of the agitation on the interaction dynamics.

The rest of this paper is organized as follows. In \S2, we present high-speed imaging of bubbles rising near the channel wall. We identify preferred configurations of pairwise bubbles.
In \S3, we introduce the model. 
In \S4, we validate and calibrate the model.
In \S5, we simulate the motions of the pairwise bubbles. We elucidate the mechanism of the preferred configurations from the perspectives of the timescales of the interaction dynamics.
In \S6, we discuss connections of the identified pairwise dynamics and the bubble cluster formation. We capture the timescale of the formation of chain-line horizontal bubble clusters in the imaging and use the model to analyze the effects of viscous interaction and the agitation on the pairwise clustering in this timescale.
In \S7 we state conclusions.
In appendix, we provide details of the model formulation.

\section{Experimental imaging}
\label{sec:exp}
\subsection{Setup}
\begin{figure}
  \subfloat[]{\includegraphics[width=6cm]{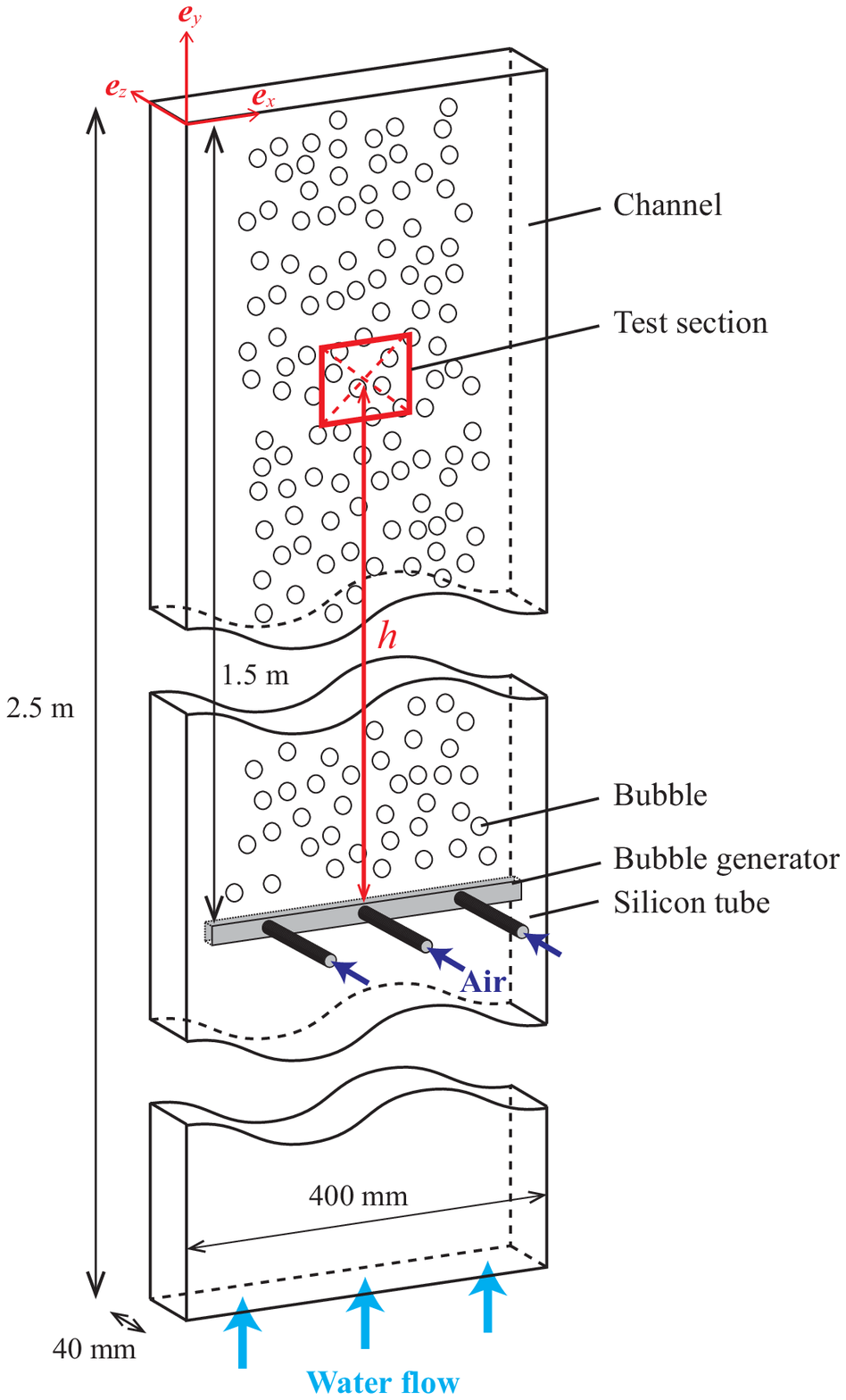}}
  \subfloat[]{\includegraphics[trim=0 -200mm 0 0,clip,width=6.8cm]{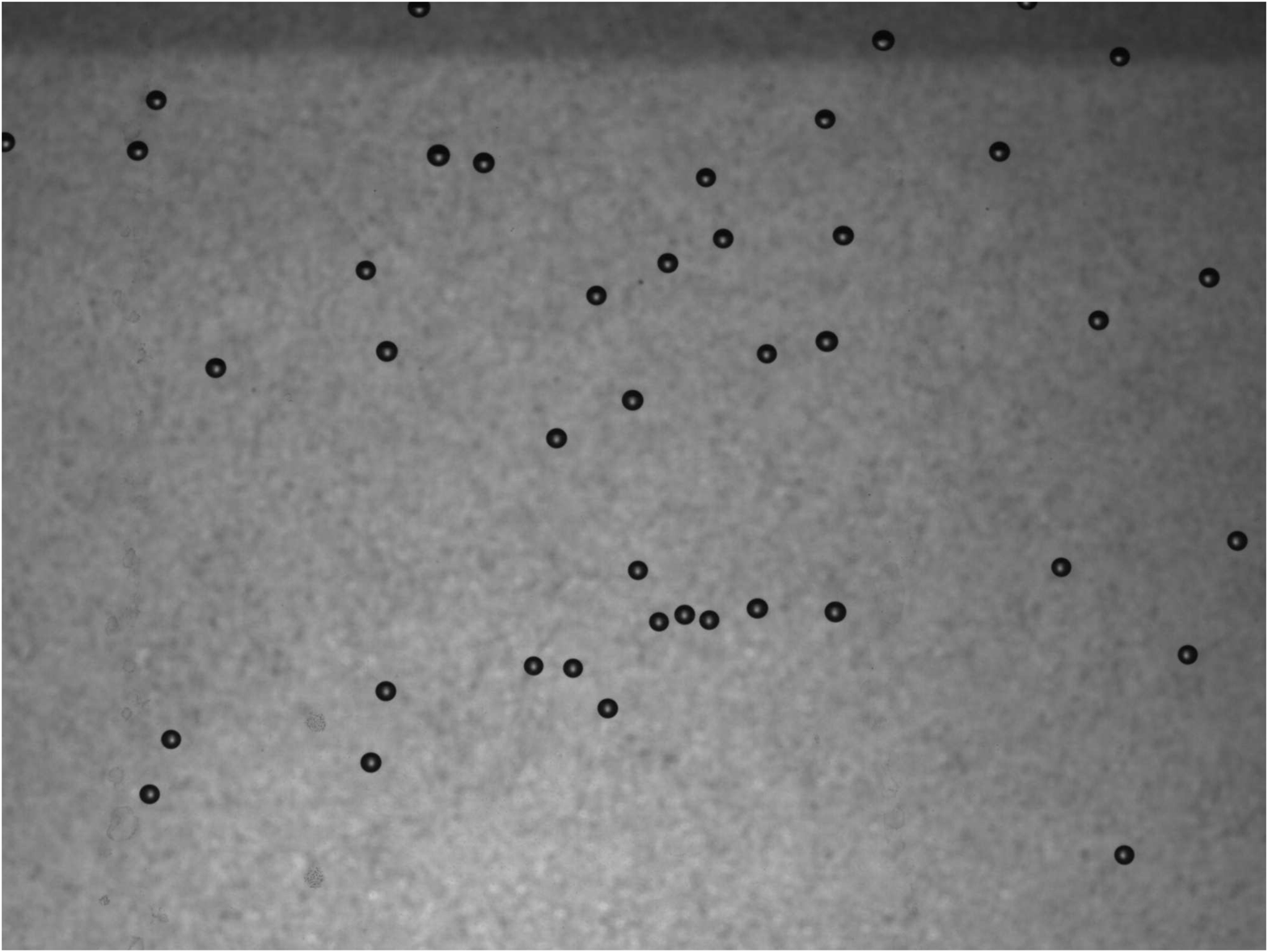}}
  \caption{(a) Schematic of the experimental setup of high-speed imaging. (b) Representative image of bubbles captured in the test section.}
\label{fig:ka1}
\end{figure}
Figure \ref{fig:ka1}a shows the schematic of the experimental setup.
The vertical channel is made of transparent acrylic plastic.
The lower end of the channel is connected to an insulator and the upper end to a tank. The insulator and the tank are connected by a plastic pipe. Water is driven upward in the vertical channel by a pump. The dimensions of the horizontal section of the channel is $40\times400$ mm and the channel height is 2.5 m.
We define the frame of reference ($\vector{e}_x$, $\vector{e}_y$, $\vector{e}_z$) such that $\vector{e}_x$ and $\vector{e}_y$ are aligned with the spanwise and vertical directions of the channel, respectively. Bubbles are released from a bubble generator attached in the channel 1.5 m above the lower end of the channel. The bubble generator is composed of 180 stainless steel pipes with 0.1 mm diameter. The pipes are vertically embedded in the wall with a uniform length of 3.0 mm above the wall. The air is supplied to the pipes by a compressor through silicon tubes. The gas flow rate is set such that monodisperse bubbles of 1.0 mm diameter are released from the pipes.
The gas flow is turned on and off by a solenoid pressure valve attached between the compressor and the tubes.
The valve is digitally programmed such that the bubbles are injected at desired timings and duty cycles.

We capture the images of bubbles rising near the wall using a high-speed camera (Photoron SA-2). The test section of the camera is square with a dimension of $60\times60$ mm and placed at the middle of the channel span and focused on the front surface of the channel. The edges of the section are parallel to the vertical and spanwise directions of the channel.
The vertical distance between the center of the test section and the bubble generator, $h$, can be varied.
Throughout the study, the camera continuously captures images with a frame rate of 1000 fps with a resolution of $2048\times2048$ pixels. A representative image is shown in figure \ref{fig:ka1}b.
The liquid flow rate is set such that the bulk Reynolds number is $Re=5100$ without gas flow. 1-Pentanol is added in water with a concentration of 20 ppm to prevent the coalescence of bubbles.

\subsection{Configuration of pairwise bubbles}
We use the setup to capture the configurations of pairwise bubbles. The test section is placed at $h=0.3$ m. To minimize the disturbance in the flow field caused by the injection of airs as well as to maintain a low void fraction, we intermittently opened the solenoid valve for a period of 0.04 s with a duty cycle 25. By this operation, sparse bubbles are periodically released in the channel without forming clusters.
From the high-speed images, we extract isolated pairs of bubbles with a condition that no bubble is present within 10 diameter from the centers of both bubbles.
The mean diameter of the sampled bubbles is 1.0 mm with a variation of 4.9\%.

\begin{figure}
  \centerline{\includegraphics[width=75mm]{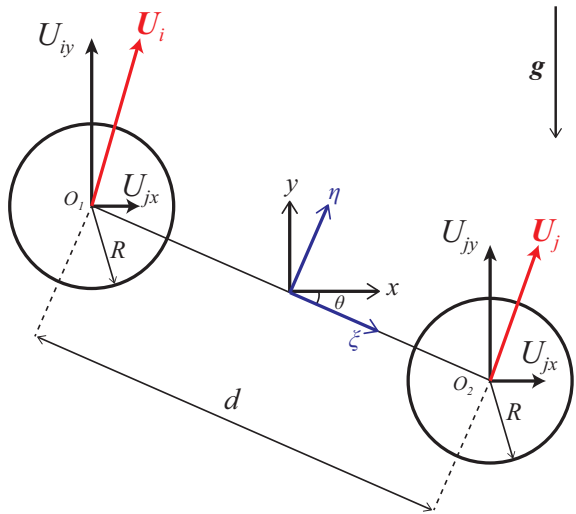}}
  \caption{Configuration of a pair of two bubbles.}
\label{fig:ka}
\end{figure}
As shown in figure \ref{fig:ka}, we define the configuration of pairwise bubbles, namely bubble-$i$ and bubble-$j$, in the $x$-$y$-$z$ Cartesian coordinates with its origin located at the middle-point of the line connecting the center of the bubbles, $O_i$ and $O_j$. The separation distance between the centers is $O_iO_j=d$. We define the non-dimensional separation distance as $S=d/R$. The pairwise angle $\theta_{ij}$ is defined between the $x$-axis and the line connecting the center of bubbles such that $-\pi/2\leq\theta_{ij}\leq\pi/2$.
Additionally we define the $\xi$-$\eta$-$z$ Cartesian coordinates, which share its origin with the $x$-$y$-$z$ Coordinates. The $\xi$-axis and the $\eta$-axis are respectively parallel and orthogonal to the line connecting the center of bubbles. The $x$-$y$-$z$ coordinates and the $\xi$-$\eta$-$z$ coordinates are associated with the two frames of reference, ($\vector{e}_x, \vector{e}_y, \vector{e}_z$) and ($\vector{e}_{\xi}, \vector{e}_{\eta}, \vector{e}_z$). $\vector{U}_i$ and $\vector{U}_j$ represent the velocities of bubble-$i$ and bubble-$j$, respectively. The gravity is taken downward along the $y$-axis as $\vector{g}=-g\vector{e}_y$. Throughout the study, we consider that the wall-normal velocity of the bubble, $U_z$, is zero.

To analyze the angle configuration of the pairwise bubbles and its dependence on the inter-bubble distance, we define the conditional pair probability distribution function (C-PPDF) as
\begin{align}
\Gamma_{C}(S, \theta)
&=
\frac{\Omega_{C}}{N_p^2}\sum_{k=1}^{N_p}\sum_{l=1}^{N_p}\delta(S-S_k)\delta(\theta-|\theta_l|)\label{eqn:C-PPDF},
\end{align}
where $N_p$ is the number of pairs sampled, $\Omega_{C}$ is normalization constant, and $H$ is the Heviside's step function.
$S_k$ and $\theta_l$ are the inter-bubble distance of the $k-$th pair and the pairwise angle of the $l$-th pair, respectively.
For convenience, we normalize C-PPDF at each $S$ to obtain the conditional angular pair distribution function (C-APDF) as
\begin{align}
G_{C}(\theta)=\frac{\Gamma_{C}(S,\theta)}{\int \Gamma_{C}(S,\theta)d\theta},
\end{align}
which corresponds to the probability of observing pairs with a pairwise angle of $\theta$, among those with an inter-bubble distance of $S$.
To obtain $\Gamma_{C}$ and $G_{C}$ from discrete bubbles, similar to an approach of \citet{Bunner02}, we approximate $\delta(S)$ and $\delta(\theta)$ by averaging over a circular strip with a thickness of $\Delta S$ with a radius of $S$, and a circular section with a central angle of $\Delta \theta$, respectively.
We use $\Delta r=1.0$ and $\Delta\theta=\pi/10$.

\begin{figure}
  \centerline{\includegraphics[width=10cm]{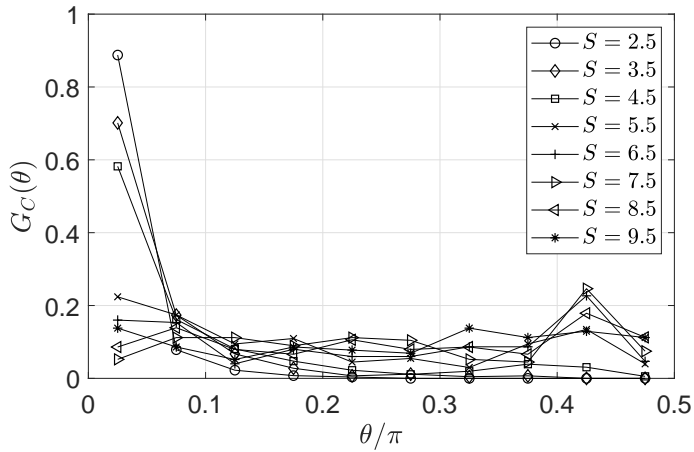}}
  \caption{Conditional angular pair distribution function for isolated pairs obtained from the experimental images.}
\label{fig:ka2}
\end{figure}
Figure \ref{fig:ka2} shows obtained C-APDF at various $S$. For pairs at $5/2\leq S<9/2$, more than half of the pairs take the pairwise angle within $0<\theta<\pi/10$. For pairs at $11/2<S<17/2$, the probability is more uniform about $\theta$, but peaks appear at $2\pi/5<\theta<9\pi/10$. Thus, pairs with a short inter-bubble distance tend to take the side-by-side configurations while those with a long inter-bubble distance tend to take nearly inline, oblique configurations. The peak decays as $S$ becomes large to approach $10$. These biases in the pairwise angle indicate that the motions of bubbles are subjected to the inter-bubble interactions.

\begin{figure}
  \centerline{\includegraphics[width=75 mm]{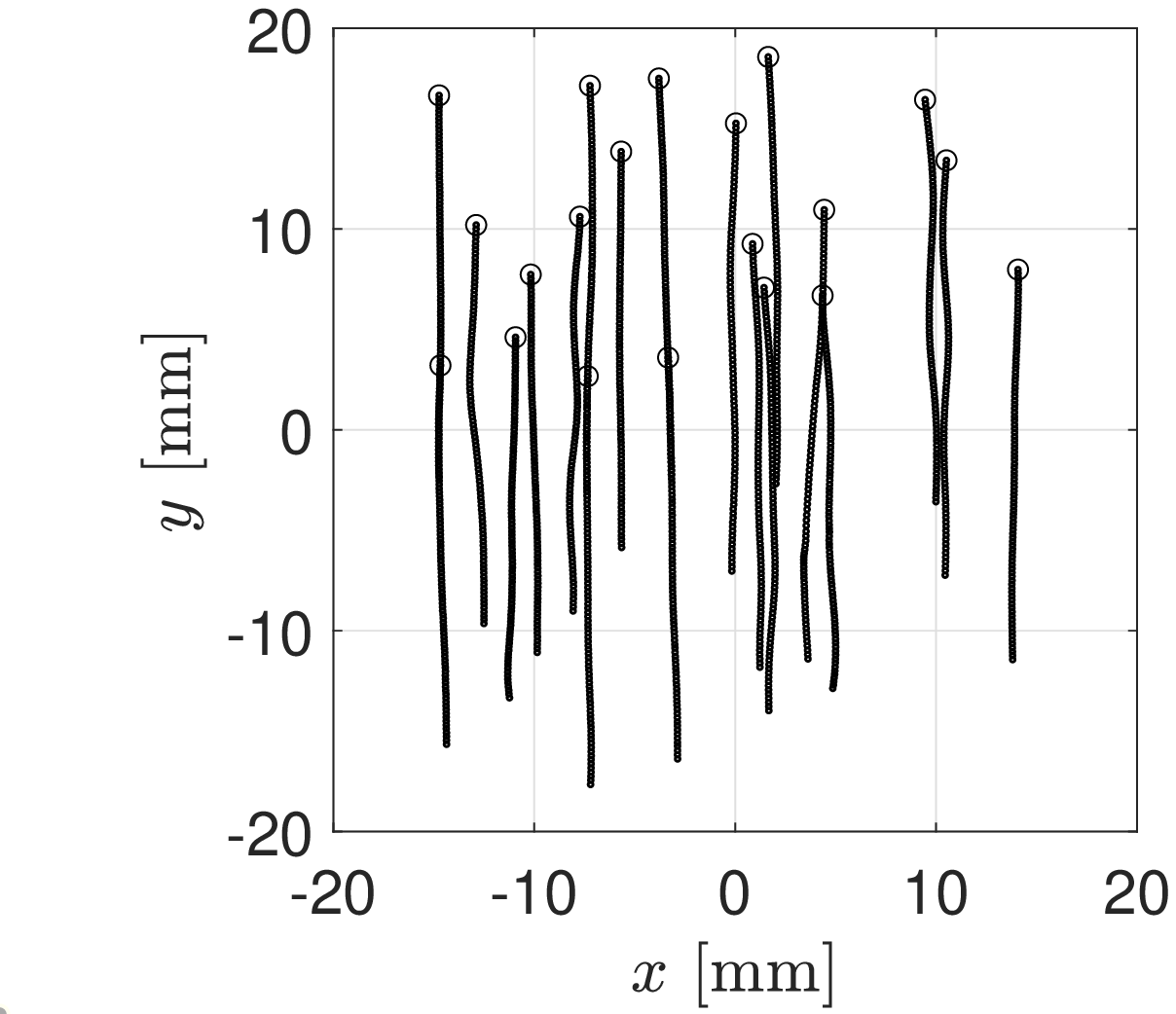}}
  \caption{Trajectories of representative bubbles captured in the test section.}
\label{fig:trajex}
\end{figure}
Figure \ref{fig:trajex} shows the trajectories of representative bubbles captured in the images.
The trajectories are not perfectly straight along $\vector{e}_y$ perpendicular to the gravity but fluctuated.
The fluctuations are observed for bubbles regardless of their proximity to others, indicating that the bubbles are subjected to turbulent agitation.

\section{Modeling}
\subsection{Formulation}
We model the translational motions of the Lagrangian point bubbles shown in figure \ref{fig:ka}.
The formulation extends that reported by \citet{Maeda13} to model the effect of turbulent agitation.
The motion of a spherical bubble immersed in a fluid can be formulated as a point-mass undergoing a translational motion \citep{Auton88, Magnaudet00, Merle05}. From this perspective, the equation of motion of an isolated spherical bubble of radius $R$ rising under buoyancy at velocity $\vector{U}$ in an incompressible viscous fluid can be expressed as
\begin{eqnarray}
(\rho_{\rm{B}}+C_{\rm{M}}\rho_{\rm{L}})\it{V}_{\rm{B}}\frac{d\vector{U}}{dt}
=
-\rm{\frac{3}{8}}\rho_{\rm{L}}\frac{\it{V}_{\rm{B}}}{\it{R}}\it{C}_{\rm{D}}|\vector{U}|\vector{U}
+(\rho_{\rm{B}-}
\rho_{\rm{L}})\it{V}_{\rm{B}}\vector{g}, \label{eqn}
\end{eqnarray}
where $\rho_{\rm B}$ and $\rho_{\rm L}$ are the densities of the bubble and the fluid respectively, $\it C_{\rm D}$ is the drag coefficient, $\it C_{\rm M}$ is the added-mass coefficient, $V_{\rm B}$ is the bubble volume. Considering that $\rho_b\ll\rho_L$ and $C_M=1/2$ \citep{Batchelor67}, the equation can be simplified as
\begin{eqnarray}
\frac{1}{2}\rho_{\rm{L}}\it{V}_{\rm{B}}\frac{d\vector{U}}{dt}
=
-\rm{\frac{3}{8}}\rho_{\rm{L}}\frac{\it{V}_{\rm{B}}}{\it{R}}\it{C}_{\rm{D}}|\vector{U}|\vector{U}
-
\rho_{\rm{L}}\it{V}_{\rm{B}}\vector{g}, \label{eqn2}
\end{eqnarray}
where the first and second terms in the right-hand side correspond to the drag and the buoyancy force, respectively.
For a spherical bubble moving in an incompressible viscous fluid, \cite{Levich62} derived $\it{C}_{\rm{D}}=Re/{\rm48}$ by considering the global kinetic energy balance in viscous potential flow.
\citet{Moore63} derived $\it{C}_{\rm{D}}=Re/{\rm48}(\rm{1}\it-M_{\infty}/Re^{\rm1/2})$, where the correction $M_{\infty}=2.211\dots$ models the effect of the viscous diffusion in the boundary layer at the bubble surface and the viscous wake formed behind the bubble formed due to the separation of the layer. For later convenience, we decompose the drag coefficient into the potential and viscous contributions as $\it{C}_{\rm{D}}=\it{C}_{\rm{Dpot}}+\it{C}_{\rm{Dvis}}$, where $\it{C}_{\rm{Dpot}}={\rm48}/Re$ and $\it{C}_{\rm{Dvis}}=M_{\infty}/{\rm48}Re^{\rm3/2}$.
For the present pairwise bubbles, we consider additional terms that represent the interaction force and turbulent agitation. The equation of motion for bubble-$i$ is expressed as
\begin{eqnarray}
\frac{1}{2}\rho_{\rm{L}}\it{V}_{\rm{B}}\frac{d\vector{U}_{i}}{dt}
=
-\rm{\frac{3}{8}}\rho_{\rm{L}}\frac{\it{V}_{\rm{B}}}{\it{R}}\it{C}_{\rm{D}}|\vector{U}_{i}|\vector{U}_{i}
-
\rho_{\rm{L}}\it{V}_{\rm{B}}\vector{g}
+
\vector{F}_{\rm{Int}\it{ij}}
+
\vector{F}_{Wi},\label{eqn:final}
\end{eqnarray}
where $\vector{F}_{\rm Int\it ij}$ represents the interaction force exerted on bubble-$i$ due to the presence of bubble-$j$, and $\vector{F}_{Wi}$ represents the fluctuation due to turbulent agitation.
We decompose $\vector{F}_{\rm int\it ij}$ into the drag and lift components as
\begin{equation}
\vector{F}_{\rm{Int}\it{ij}}=\vector{F}_{\rm{D_{Int}}\it{ij}}+\vector{F}_{\rm{L_{Int}}\it{ij}},
\end{equation}
We further decompose $\vector{F}_{\rm{D_{Int}}\it{ij}}$ and $\vector{F}_{\rm{L_{Int}}\it{ij}}$ into the contribution of the potential flow (potential interaction) and that of the viscosity (viscous interaction) as
\begin{equation}
\vector{F}_{\rm{D_{Int}}\it{ij}}=\vector{F}_{\rm{D_{Int}}\it{ij}\rm pot}+\vector{F}_{\rm{D_{Int}}\it{ij}\rm vis},
\end{equation}
\begin{equation}
\vector{F}_{\rm{L_{Int}}\it{ij}}=\vector{F}_{\rm{L_{Int}}\it{ij}\rm pot}+\vector{F}_{\rm{L_{Int}}\it{ij}\rm vis}.
\end{equation}

We normalize each term on the right-hand-sides of these equations by $1/2\rho |\vector{U}_i|^2\pi R^2$ to obtain the corresponding potential and viscous contributions of the drag and lift coefficients as ${C}_{\rm{D_{Int}}\it{ij}\rm pot}$, ${C}_{\rm{L_{Int}}\it{ij}\rm pot}$, ${C}_{\rm{D_{Int}}\it{ij}\rm vis}$, and ${C}_{\rm{L_{Int}}\it{ij}\rm vis}$.

\subsection{Potential interaction}
We denote the velocity potential of fluid surrounding the bubbles as $\phi$. To induce ${C}_{\rm{D_{Int}}\it{ij}\rm pot}$ and ${C}_{\rm{L_{Int}}\it{ij}\rm pot}$, we employ a bipoler decomposition of $\phi$ on the ($\xi$, $\eta$, $z$) coordinate \citep{Wijngaarden76,Kok93}:
\begin{eqnarray}
\phi=\phi_i+\phi_j,
\end{eqnarray}
where $\phi_i$ and $\phi_j$ are the velocity potentials expanded from $O_i$ and $O_j$, respectively. We define the spherical coordinate systems, $r_i$-$\psi_i$-$\varphi$ coordinates and $r_j$-$\psi_j$-$\varphi$ coordinates, with their origins located on $O_i$ and $O_j$, respectively. Using Legendre polynomials and associated Legendre polynomials, $\phi_i$ and $\phi_j$ can be expressed as
\begin{eqnarray}
\phi_i&=&\sum_{n=1}^{\infty}(\frac{R}{r_i})^{n+1}\{g^{(1)}_{in}P_n(\cos{\psi_i})+(g^{(2)}_{in}\cos{\varphi}+g^{(3)}_{in}\sin{\varphi})P^1_n(\cos{\psi_i})\}, \\
\phi_j&=&\sum_{n=1}^{\infty}(\frac{R}{r_j})^{n+1}\{g^{(1)}_{jn}P_n(\cos{\psi_j})+(g^{(2)}_{jn}\cos{\varphi}+g^{(3)}_{jn}\sin{\psi})P^1_n(\cos{\psi_j)}\},
\end{eqnarray}
where $g^{(k)}_{in}$, and $g^{(k)}_{jn}$ ($k=1, 2, 3$) are coefficients determined by the boundary condition at the surface of the bubbles.
For convenience, we decompose the interaction force $\vector{F}_{\rm{Int}\it{ij}\rm{pot}}$ into components corresponding to the forces along the $\xi$- and $\eta$- axes as $\vector{F}_{\rm{Int}\it{ij}\rm{pot}}={\vector{F}}_{\xi_{\rm Int}ij \rm pot}+{\vector{F}}_{\eta_{\rm Int}ij \rm pot}$. ${\vector{F}}_{\xi_{\rm Int}ij \rm pot}$ and ${\vector{F}}_{\eta_{\rm Int}ij \rm pot}$ are expressed as polynomials in terms of $S$:
\begin{eqnarray}
{\vector{F}}_{\xi_{\rm Int}ij \rm pot}&=&f_{\xi_{\rm Int}ij\rm pot}^{(-4)}S^{-4}+f_{\xi_{\rm Int}ij\rm pot}^{(-7)}S^{-7}+\cdots, \\
{\vector{F}}_{{\eta_{\rm Int}ij \rm pot}}&=&f_{\eta_{\rm Int}ij\rm pot}^{(-4)}S^{-4}+f_{\eta_{\rm Int}ij\rm pot}^{(-7)}S^{-7}+\cdots,
\end{eqnarray}
where
\begin{eqnarray}
f_{\xi_{\rm Int}ij\rm pot}^{(-4)}&=&\rho\pi R^2({-6{U}^{2}_{j\xi}+3{U}_{i\eta}{U}_{j\eta}}), \\
f_{\xi_{\rm Int}ij\rm pot}^{(-7)}&=&\rho\pi R^2 (-6U^2_{i\xi}+12U_{i\xi}U_{j\xi}+6U^2_{j\xi}+\frac{3}{2}{U^2_{i\eta}}+\frac{3}{2}{U^2_{j\eta}}),\\
f_{\eta_{\rm Int}ij\rm pot}^{(-4)}&=&\rho\pi R^2 3{U}_{j\xi}({U}_{i\eta}+{U}_{j\eta}), \\
f_{\eta_{\rm Int}ij\rm pot}^{(-7)}&=&\rho\pi R^2\{(-\frac{9}{2}U_{i\xi}+3U_{j\xi})U_{i\eta}-\frac{3}{2}U_{j\xi}U_{j\eta}\}.
\end{eqnarray}
Normalizing ${\vector{F}}_{\xi_{\rm Int}ij \rm pot}$ and ${\vector{F}}_{\xi_{\rm Int}ij \rm pot}$ by $1/2\rho|\vector{U}_i|^2\pi R^2$, we define the dimensionless coefficients ${C}_{\xi_{\rm Int}ij \rm pot}$ and ${C}_{\eta_{\rm Int}ij \rm pot}$ expressed as
\begin{eqnarray}
{C}_{\xi_{\rm Int}ij \rm pot}&=&c_{\xi_{\rm Int}ij\rm pot}^{(-4)}S^{-4}+c_{\xi_{\rm Int}ij\rm pot}^{(-7)}S^{-7}+\cdots, \\
{C}_{{\eta_{\rm Int}ij \rm pot}}&=&c_{\eta_{\rm Int}ij\rm pot}^{(-4)}S^{-4}+c_{\eta_{\rm Int}ij\rm pot}^{(-7)}S^{-7}+\cdots,
\end{eqnarray}
where
\begin{eqnarray}
c_{\xi_{\rm Int}ij\rm pot}^{(-4)}&=&\frac{2f_{\xi_{\rm Int}ij\rm pot}^{(-4)}}{\rho\pi R^2|\vector{U}_i|^2}=\frac{-12{U}^{2}_{j\xi}+6{U}_{i\eta}{U}_{j\eta}}{|\vector{U}_{i}|^2}, \label{cgijpot4}\\
c_{\xi_{\rm Int}ij\rm pot}^{(-7)}&=&\frac{2f_{\xi_{\rm Int}ij\rm pot}^{(-7)}}{\rho\pi R^2|\vector{U}_i|^2}=\frac{-12U^2_{i\xi}+24U_{i\xi}U_{j\xi}+12U^2_{j\xi}+3{U^2_{i\eta}}+3{U^2_{j\eta}}}{|\vector{U}_{i}|^2},\label{cgijpot7}\\
c_{\eta_{\rm Int}ij\rm pot}^{(-4)}&=&\frac{2f_{\eta_{\rm Int}ij\rm pot}^{(-4)}}{\rho\pi R^2|\vector{U}_i|^2}=\frac{6{U}_{j\xi}({U}_{i\eta}+{U}_{j\eta})}{|\vector{U}_{i}|^2}, \label{ceijpot4}\\
c_{\eta_{\rm Int}ij\rm pot}^{(-7)}&=&\frac{2f_{\eta_{\rm Int}ij\rm pot}^{(-7)}}{\rho\pi R^2|\vector{U}_i|^2}=\frac{(-9U_{i\xi}+6U_{j\xi})U_{i\eta}-3U_{j\xi}U_{j\eta}}{|\vector{U}_{i}|^2}.\label{ceijpot7}
\end{eqnarray}
Using the relations between the $x$-$y$-$z$ coordinates and the $\xi$-$\eta$-$z$ coordinates, we obtain
\begin{eqnarray}
{C}_{D_{\rm Int}ij \rm pot}&=&-{C}_{\xi_{\rm Int}ij \rm pot}\cos{\theta}+{C}_{\eta_{\rm Int}ij \rm pot}\sin{\theta}, \\
{C}_{L_{\rm Int}ij \rm pot}&=&-{C}_{\xi_{\rm Int}ij \rm pot}\sin{\theta}+ {C}_{\eta_{\rm Int}ij \rm pot}\cos{\theta}.
\end{eqnarray}
Further details of the formulations are provided in Appendix A.

\subsection{Viscous interaction}
To express the viscous contributions of the lift and drag forces due to the inter-bubble interaction, ${\vector{F}}_{D_{\rm Int}\it ij \rm vis}$ and ${\vector{F}}_{L_{\rm Int}ij \rm vis}$, we follow an analytical model introduced by Hallez and Legendre (2011) (hereafter we will recall as HLmodel). HLmodel describes forces acting on pairwise spherical bubbles under mutual interactions rising at the equal velocity at $50\leq Re\leq500$ in an incompressible viscous fluid quiescent at infinity. Particularly, from the model we extract the viscous contributions of the interaction forces induced by the wake formed behind the leading bubble on the trailing bubble as well as those induced by the viscous diffusion in the boundary layers at the surface of bubbles. 
Although the velocities of the two bubbles in our model are not necessarily equal, they are rising upward under buoyancy at similar velocities: $\vector{U}_i\simeq\vector{U}_j\simeq{U}_{iy}\vector{e}_y$.
Based on the similarity, we adopt the HL model to express ${\vector{F}}_{D_{\rm Int}\it ij \rm vis}$ and ${\vector{F}}_{L_{\rm Int}ij \rm vis}$ as
\begin{eqnarray}
\vector{F}_{\rm{D_{Int}}\it{ij}\rm vis}&\approx&-\rm{\frac{3}{8}}\rho_{\rm{L}}\frac{\it{V}_{\rm{B}}}{\it{R}}\it{C}_{\rm{D}_{Int}\it{ij}\rm vis}|\vector{U}_{\it{i}}|^{\rm2}\vector{e}_{y}, \\
\vector{F}_{\rm{L_{Int}}\it{ij}\rm vis}&\approx&\rm{\frac{3}{8}}\rho_{\rm{L}}\frac{\it{V}_{\rm{B}}}{\it{R}}\it{C}_{\rm{L}_{Int}\it{ij}\rm vis}|\vector{U}_{\it{i}}|^{\rm2}\vector{e}_{x},
\end{eqnarray}
where $C_{\rm{D_{Int}\it ij\rm vis}}$ and $C_{\rm{D_{Int}\it ij\rm vis}}$ are the viscous contributions of the drag and lift coefficient of bubble-$i$ in the presence of bubble-$j$:
\begin{equation}
{C}_{\rm{D}_{Int}\it{ij}\rm vis}=\left\{
\begin{array}{ll}
\displaystyle
-M_{\infty}\frac{48}{Re^{\frac{3}{2}}}S^{-3}, &0\leq\theta\leq\pi,\\[12pt]
\displaystyle
-(M^{*}_{2}(1+S^3)-M_{\infty})\frac{48}{Re^{\frac{3}{2}}},&-\pi\leq\theta\leq0.\label{cdvis}
\end{array} \right.
\end{equation}
\begin{equation}
C_{\rm{L_{Int}\it ij\rm vis}}=\left\{
\begin{array}{ll}
\displaystyle
-\frac{240}{ReS^{4}}\cos{\theta}, &0\leq\theta\leq\pi,\\[12pt]
\displaystyle
-\frac{240}{ReS^{4}}\cos{\theta}-\frac{2}{3}C_{\rm{L\,wake}}\beta Re\frac{\cos{\theta}}{\sin{\theta}}(1-\tilde{u}_{\rm{wake}})\tilde{u}_{\rm{rot}}, &-\pi\leq\theta\leq0.\label{Clvis}
\end{array} \right.
\end{equation}
When the pairwise angle is at $0\leq\theta\leq\pi$, bubble-$i$ is at the leading position against bubble-$j$ along the $y$-axis, and viscous interaction for bubble-$i$ comes only from the viscous diffusion in the boundary layer of bubble-$j$. On the other hand, when $-\pi\leq\theta\leq0$, bubble-$i$ is located at the trailing position against bubble-$j$ and subjected also to the influence of the wake formed behind bubble-$j$. In the wake region, dragged by the leading bubble, the distribution of the magnitude of the fluid's velocity along the direction of the bubble's motion takes a Gaussian profile which decays asymptotically to zero in regions distant from the wake \citep{Batchelor67}. As discussed in detail by \citet{Yuan94, Harper97}, due to this local velocity distribution within the wake region, the drag coefficient of the trailing bubble in the wake region against the background flow field is effectively reduced (slip-stream effect). The coefficient $M_2^{*}$ in relation (\ref{cdvis}) accounts for this drag reduction. Moreover, due to the non-zero vorticity distribution in the wake, the trailing bubble is subjected to the shear-induced lift force in the direction outward from the wake. The magnitude of this shear-induced lift force is expressed as $C_{L_{\rm Single}} \vector{u}\times(\vector{\nabla\times\vector{u}})$, where $C_{L_{\rm Single}}$ is the lift coefficient of a single isolated bubble immersed in an uniform shear flow. \citet{Auton87} analytically induced $C_{L_{\rm Single}}=1/2$ in the limit of weak shear. \citet{Legendre98} extended the result as $C_{L_{\rm Single}}=(Re+16/2(Re+29))$ for $Re\geq5$. HLmodel modifies this expression to define the lift, which appears in the term that includes $C_{\rm{L\,wake}}$.
This wake-induced lift force plays a key role in the pairwise dynamics considered in the present study.
In the following sections, in order to assess the effects of the viscous contribution on the interaction force, we simulate cases without including viscous interactions by setting ${C}_{\rm{D}_{Int}\it{ij}\rm vis}={C}_{\rm{L}_{Int}\it{ij}\rm vis}=0$, and compare results obtained with the complete model.

\subsection{Turbulent agitation}
The random motions of particles in turbulent flow field can be modeled as the Brownian motion \citep{Taylor22}.
Meanwhile, the fluctuations of the velocities of the present pairwise bubbles with a finite inter-bubble distance may be correlated, since the turbulent agitation that drives the fluctuations may have spatial structures at the scale of the inter-bubble distance.
For pairwise particles that move on the stream lines of the background flow field (tracer-particles), their velocity correlation corresponds to the Lagrangian velocity correlation of the background single-phase turbulence \citep{Batchelor53,Durbin80,Thomson90,Sawford01}.
On the contrary, the present bubbles have relative velocities (slip velocity) against the career fluid, and the resulting trajectories do not follow the stream lines of the fluid that would otherwise be present at the space filled by the bubbles.
The correlation of the fluctuating velocities of bubbles therefore may not correspond to the Lagrangian velocity correlation of single phase turbulence.
Systematic derivation of such a correlation is not a simple task.
The slip velocity results from the inter-phase momentum transfer at the bubble interface, and computation of this momentum transfer requires modeling the thin boundary layer at the surface of bubbles. The state of the boundary layer can be influenced by the surfacant molecules as well as dynamically change due to interactions with turbulent eddies in the small scale. The consideration of such a complex dynamics is a subject of high-fidelity simulation and, as aforementioned, out of the scope of the present study. 

Instead, we model the effect of the agitation through spatio-temporal stochastic forcing of bubbles, $\vector{F}_{Wi}$, based on an ansatz that it is Gaussian-in-time and colored-in-space.
We also crudely assume that the spatial correlation is isotropic on the $x-y$ plane.
As a result, the spatial component of the correlation can be represented by a scalar function about the inter-bubble distance.
Rather than treating the equation (\ref{eqn:final}) as a stochastic differential equation, following \citet{Maeda20}, we empirically express $\vector{F}_{Wi}$ as a pseudo-stochastic process:
\begin{equation}
    \vector{F}_{W_i} \approx \sum^{Ns}_k C_{k}\vector{\chi}_k(\mathrm{sign}(i-j)d_{ij})\sum^{Nt}_l D_{k,l}\zeta_{k,l}(t),\label{eqn:noise}
\end{equation}
where $\vector{\chi}$ and $\zeta$ are the spatial and temporal basis functions, respectively.
$C$ and $D$ are weighting coefficients that respectively define the spatial and temporal spectra of the noise.
In the present study, we employ phase-randomized Fourier basis functions for both $\vector{\xi}$ and $\zeta$.
This pseudo-stochastic representation of the agitation essentially extends the expression of \citet{Kraichnan70} that was used to represent the random velocity field in isotropic turbulence.
In the present study, the parameters are obtained such that simulations reproduce the statistics in the experiment a-posteriori, including the velocity power spectral density (PSD) and the mean-square-displacement (MSD) of isolated bubbles, as well as the spatial correlation of the fluctuating velocities of pairwise bubbles.
$(N_s, N_t)=(10^3, 10^2)$ is found sufficient to obtain statistical convergence.
The comparisons of the experimental and numerical statistics are discussed in the following section.
Further details on the stochastic representation and the parameters are provided in Appendix B.
Throughout the study, equation (\ref{eqn:final}) is integrated to simulate the motion of bubbles, using the standard fourth-order Runge-Kutta scheme (RK4) with a sufficiently small time step size.

\section{Statistics of the motions of isolated bubbles and pairwise bubbles}
\subsection{Single isolated bubble}
For isolated bubbles, the turbulent agitation is independent of the spatial coordinates of bubbles, within the assumption that the response of bubbles to the agitation is spatially homogeneous on the wall.
To characterize the fluctuating velocity from the experimental data as well as to validate the temporal component of the stochastic forcing, we address the temporal statistics of single isolated bubbles.

\begin{figure}
\centering
  \includegraphics[width=67mm,trim=0 0 0 0, clip]{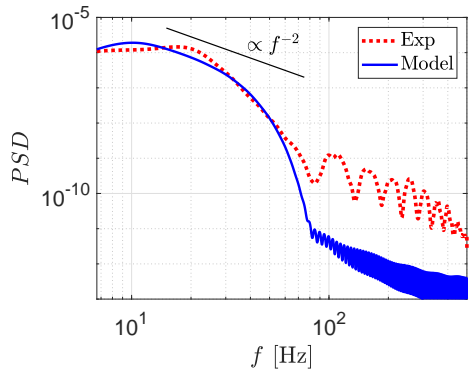}
  \caption{Ensemble averaged power spectral density of the velocity fluctuation of isolated bubbles.}
\label{fig:vevol}
\end{figure}
Figure \ref{fig:vevol} compares the ensemble averaged PSD of the fluctuating velocity of isolated bubbles obtained from the experiment and the simulation. The fluctuating velocity is defined as the mean-subtracted velocity of each bubble within a temporal window of 0.15 s.
The overall profile of the modeled PSD agrees with the experimental one, although the experimental PSD is less smooth.
The PSDs present a sharp cut-off at $f_c\approx 50$ Hz, which corresponds to the maximum frequency of the agitation.

\begin{figure}
\centering
  \includegraphics[width=67mm,trim=0 0 0 0, clip]{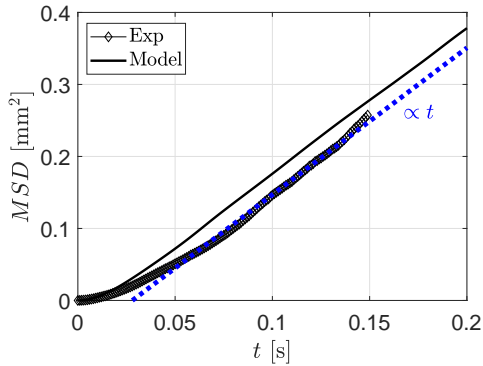}
  \caption{Ensemble averaged mean square spanwise displacement of isolated bubbles.}
\label{fig:msd}
\end{figure}
Figure \ref{fig:msd} compare the spanwise MSDs of the same isolated bubbles obtained in the experiment and the simulation.
The overall profiles of MSD agree with each other, except that the modeled MSD is slightly shifted upward due to the faster initial growth.
For both plots, MSD presents a faster growth near the origin.
The slope then becomes linear at $t\approx0.05$ s.
The profiles of MSD agree with the theory of turbulent diffusion \citep{Taylor22}. The initial growth corresponds to the ballistic translation of the bubbles at the time scale below that of the agitation, $1/f_c\approx 0.02$ s. The linear growth at $t>1/f_c$ corresponds to the Brownian diffusion.
The initial discrepancy between the two plots can be due to the fluctuation of the (mean) drag coefficient against the sudden acceleration, which is not formulated in the present model.
Nevertheless, the slope of the linear regime agrees very well with the experiment, and the difference between the two profiles is constant over time.
For the purpose of reduced order modeling in the present study, we consider that the difference is admissible for the aimed accuracy of the model prediction.

\begin{figure}
\centering
  \subfloat[]{\includegraphics[width=67mm,trim=0 0 0 0, clip]{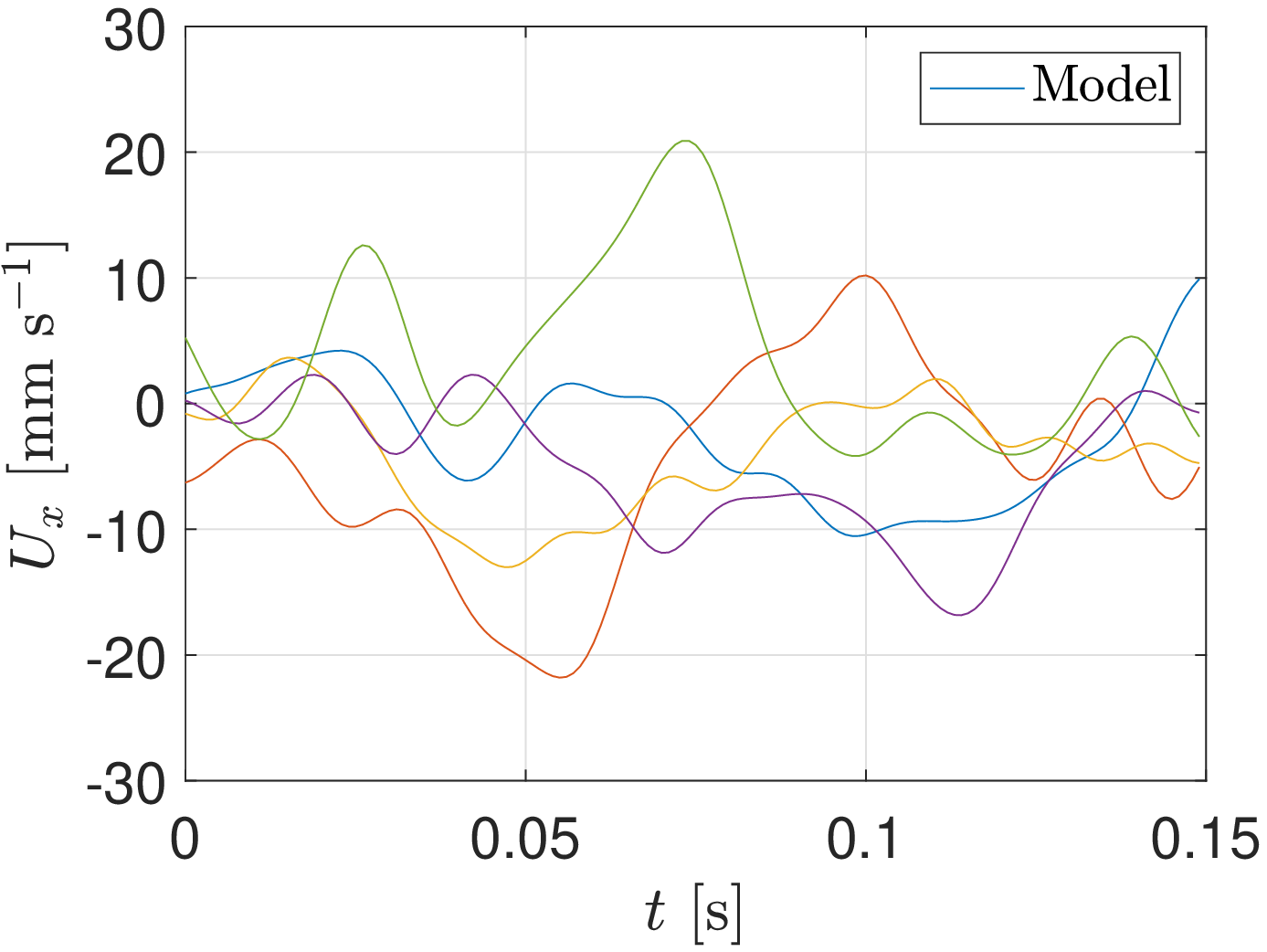}}
  \subfloat[]{\includegraphics[width=67mm,trim=0 0 0 0, clip]{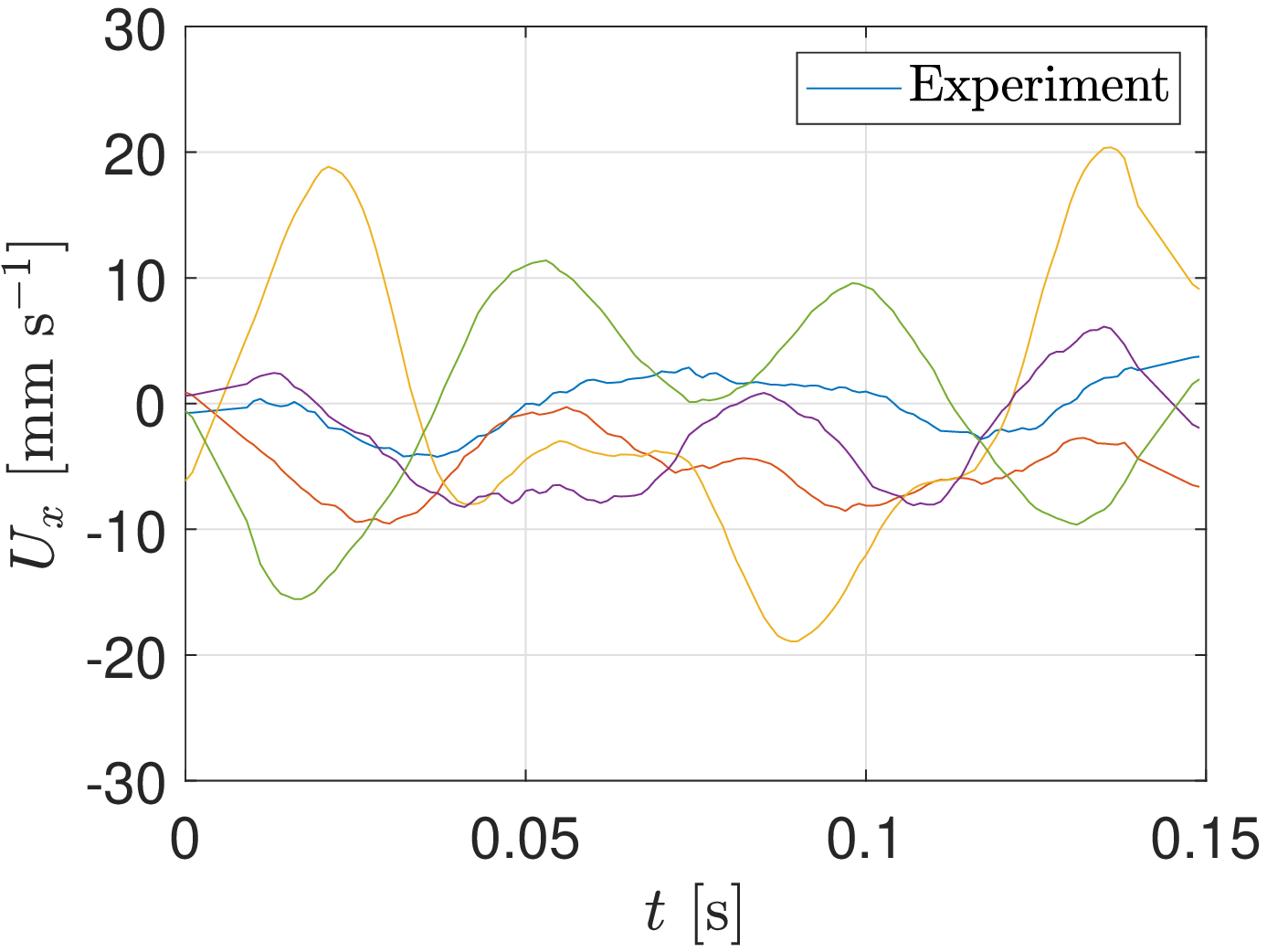}}
  \caption{Evolution of the span-wise velocity of representative bubbles in (a) experiment and (b) simulation, within the sampling time window.}
\label{fig:vcomp}
\end{figure}
Figure \ref{fig:vcomp}a and b respectively show the evolution of the fluctuating velocity of five representative bubbles which were randomly chosen from the experiment and that from the simulation. The profiles of the two figures qualitatively agree well, indicating that the modeled velocity fluctuation reproduces the experimental one.

\begin{figure}
\centering
  \subfloat[]{\includegraphics[width=67mm,trim=0 0 0 0, clip]{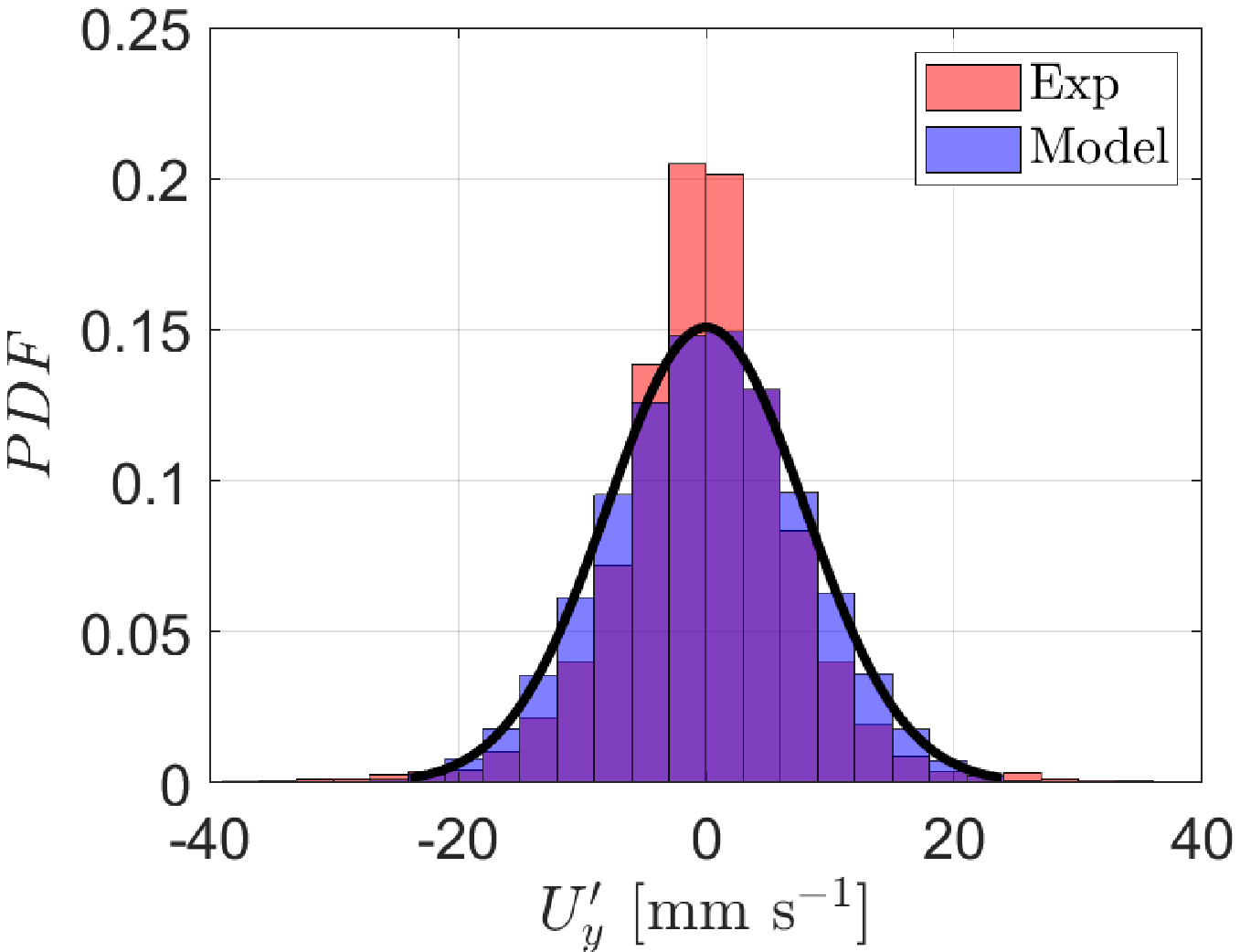}}
  \subfloat[]{\includegraphics[width=67mm,trim=0 0 0 0, clip]{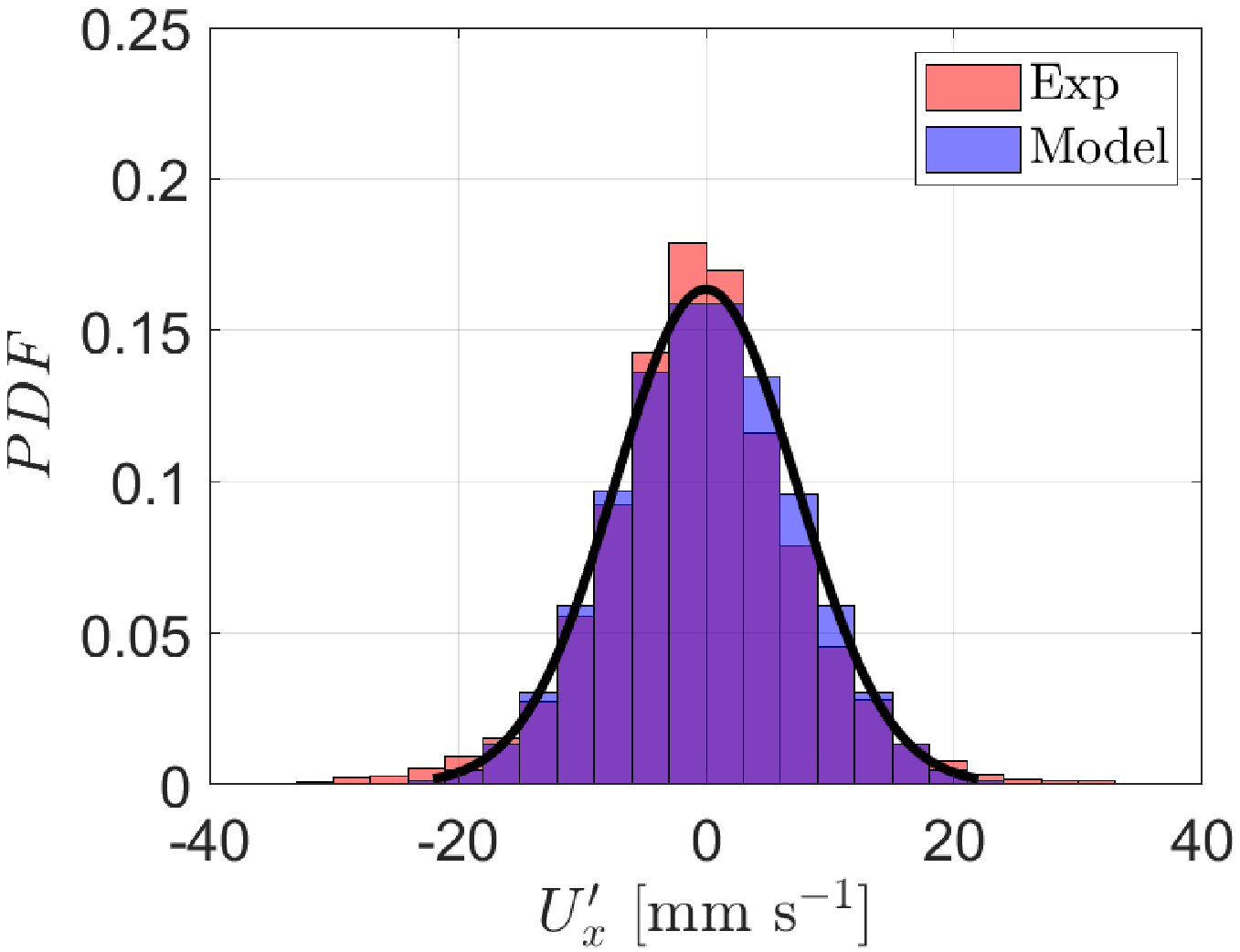}}
  \caption{Probability density functions of the (a) span-wise velocity and (b) stream-wise velocity of a single isolated bubbles.}
\label{fig:PDF}
\end{figure}
Figure \ref{fig:PDF}a and b respectively compare the PDFs of fluctuating components of the spanwise and streamwise velocities obtained from the model and the experiment.
The modeled and experimental PDFs agree with each other in both figures.
As expected from the modeling anzatz that the stochastic forcing is white-in-time, the modeled PDFs are excellently fitted by normal distribution.
Meanwhile, the experimental PDFs are non-Gaussian; their peaks at $v=0$ are greater and tails are heavier, compared to the modeled PDFs.
The heavy tail may be associated with the intermittency of turbulent fluctuations in the career fluid.
These high-kurtosis features are more pronounced in the PDF of the streamwise velocity.
The difference in the profiles of the span-wise and streamwise PDFs indicates the spatial anisotropy of the agitation, which is not considered in the present model.
Nevertheless, similar to the argument made about MSD, we consider that the discrepancies between the modeled and experimental PDFs are admissible for the purpose of the present study.

\subsection{Spatial correlation of the agitation}
In order to obtain the spatial correlation of the agitation, we compute the Lagrangian two-point correlation of the fluctuating velocities of pairs of bubbles.
To isolate the effect of potential interaction on the correlation, we exclude pairs with their inter-bubble distance smaller than 5$S$ at any instance on their trajectories.

\begin{figure}
\centering
  \includegraphics[width=67mm,trim=0 0 0 0, clip]{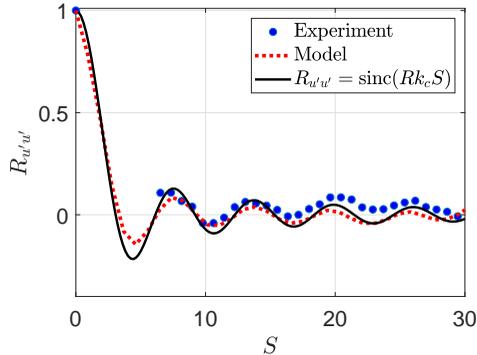}
  \caption{Second-order correlation function of the fluctuating velocities of pairwise bubbles, $R_{u'u'}$.}
\label{fig:R}
\end{figure}
Figure \ref{fig:R} shows the measured and modeled correlations, $R{u'u'}=(R_{U_x'U_x'}+R_{U_y'U_y'})/2$ as well as a fit with a sinc function, $\mathrm{sinc}(Rk_cS)$.
Interestingly, the obtained correlation shows reasonable agreement with the sinc function with $k_c=2.05$ Hz$^{-1}$rad.
Although other candidate functions can be use to fit to the data, this agreement suggests that the spectrum of the spatial correlation of the data can simply be approximated by a top-hat function with a cut of wave number of $k_c$.
The cut-off wave number indicates that the minimum spatial length-scale of the correlation is
$2\pi/k_c\approx 3$ mm.
It should again be emphasized that this integral length scale is empirical and unique to the fluctuating velocities of the bubbles and not immediately associated with the spatial correlation of the background (single-phase) turbulent flow field.
In the meantime, from an a-priori point of view, the motion of bubbles are expected to be insensitive to the velocity fluctuations whose structural length scale is on the order of or smaller than the bubble size, $O(1)$ mm.
The agreement of the a-priori and a-posteriori integral length scales of the correlation provides for additional confidence in the validity of the present model.

\section{Dynamics of the pairwise bubbles}\label{sec:rules_submission}
\subsection{Effect of viscous wake}
In order to assess the preferred configurations captured in the experiment, we simulate the motions of pairwise bubbles.
We first address pairs in initially near in-line configurations without agitation. We track the motion of the trailing bubble on coordinates whose origin is placed at the center of the trailing bubble, $(\Delta x, \Delta y) =(x_j-x, y_j-y)$. The trailing bubble is initially released at $(\Delta x_0, \Delta y_0):\Delta x_0\in[\varepsilon, 2], \Delta y_0\in[-5R, 0]$.
We terminate simulations when the bubbles collide. We set the initial velocity of both bubbles as $(U_x,U_y)=(0,0.311)$ ms$^{-1}$. The corresponding Reynolds number is $Re$=311. This velocity is the terminal rise velocity of an isolated single spherical bubble with a diameter of 1.0 mm in quiescent water, based on equation (\ref{eqn:final}).
We simulate cases with and without viscous interaction, and compare results.

\begin{figure}
  \subfloat[]{\includegraphics[width=45mm,trim=40 0 45 0, clip]{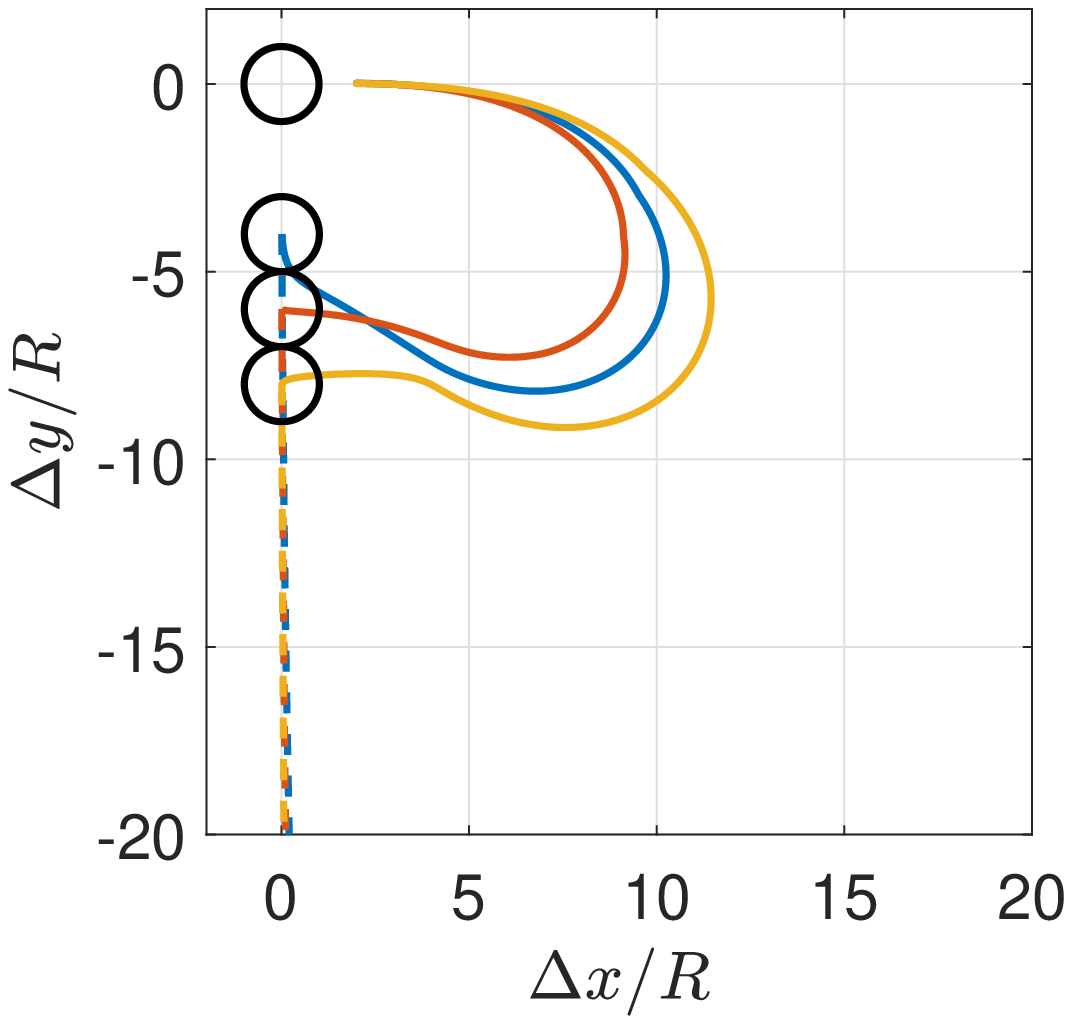}}
  \subfloat[]{\includegraphics[width=45mm,trim=40 0 45 0, clip]{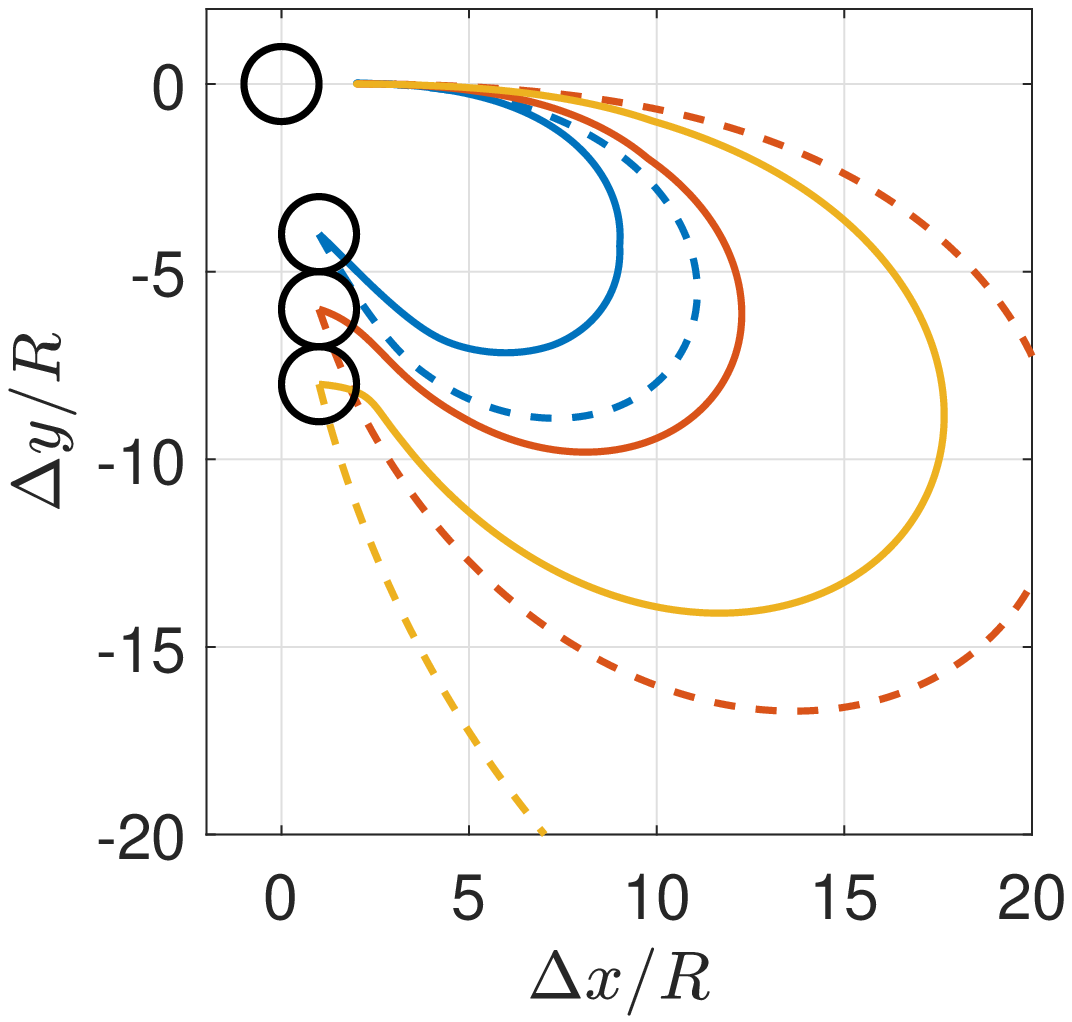}}
  \subfloat[]{\includegraphics[width=45mm,trim=40 0 45 0, clip]{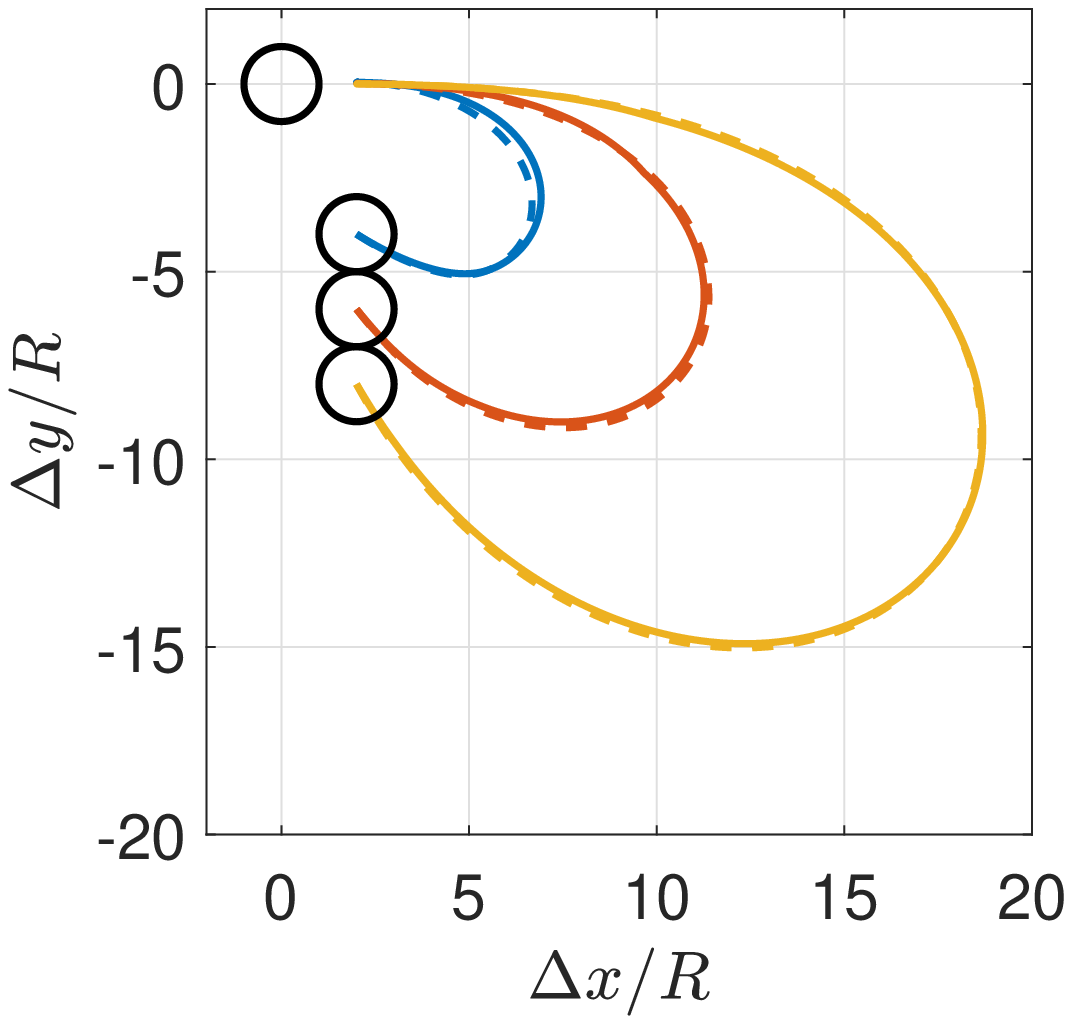}}
  \caption{Trajectories of the trailing bubble relative to the leading bubble, with various initial configurations without agitation. Results ($-$) with and ($--$) without viscous interaction are compared.}
\label{fig:traj}
\end{figure}
Figure \ref{fig:traj}a-c show the relative trajectories of the trailing bubble for representative pairs with various initial configurations.
In figure \ref{fig:traj}a, cases with $\Delta x_0=\varepsilon$ are shown in that the bubbles are initially in-line with the small perturbation. Without viscous interaction, in all the cases the trailing bubble travels away from the leading bubble, nearly vertically due to the repulsion induced by potential flow.
With viscous interaction, in the case with $\Delta y_0=4R$, the trailing bubble first travels downward and then translates in the positive $x$ direction. The bubble subsequently trails a curve-line path to horizontally approach the leading bubble. 
In the cases with $\Delta y_0=6R$ and $8R$, the trailing bubble horizontally translates in the positive $x$ direction, and then draws a similar curve-line path to approach the leading bubble.
The initial downward motion of the trailing bubble with $\Delta y_0=4R$ indicates the dominance of the potential-induced repulsion.
Figure \ref{fig:traj}b shows cases with $\Delta x_0=R$.
Without viscous interaction, for all cases, the trailing bubble first travels diagonally downward to increase the inter-bubble distance, and then draws curve-line paths to approach the leading bubble. The radius of the curve-line path increases with the initial inter-bubble distance, unlike those in figure \ref{fig:traj}a.
With viscous interaction, the bubble initially takes a more horizontal path and then trails the curve-line path.
For the same initial condition, the radius of the curve-line path becomes smaller with viscous interaction.
Compared to the cases shown in figure \ref{fig:traj}a, the difference in the trajectories due to viscous interaction is less pronounced.
Figure \ref{fig:traj}c shows cases with $x_0=2R$.
All trajectories of the trailing bubble are similar to those in \ref{fig:traj}b; the bubbles move diagonally downward and then trail curve-line paths to approach the leading bubble.
Viscous interaction makes almost no difference in this figure.

Overall, the relative trajectories shown in figure \ref{fig:traj} indicate the strong influence of the viscous wake on the pairwise dynamics. For in-line pairs in that the trailing bubble is initially at $\Delta x<2R$, the trailing bubble is pushed out from the wake region and takes significantly shorter paths to reach side-by-side with the leading bubble. The results also support the theory that the in-line configuration is unstable, as predicted by \citet{Harper70} and \citet{Hallez11}.
Note that when the pair is initially in perfectly in-line configurations ($\Delta x_0 = 0$), the bubbles reach its equilibrium state such that $(\Delta x, \Delta y)=(0, -6.8R)$, with a Reynolds number of $Re_{eq}=326$. The equilibrium distance and $Re_{eq}$ correspond to those derived in previous studies (\citet{Yuan94, Harper97, Hallez11}).

\begin{figure}
\centering
  \subfloat[]{\includegraphics[width=55mm,trim=25 0 45 0, clip]{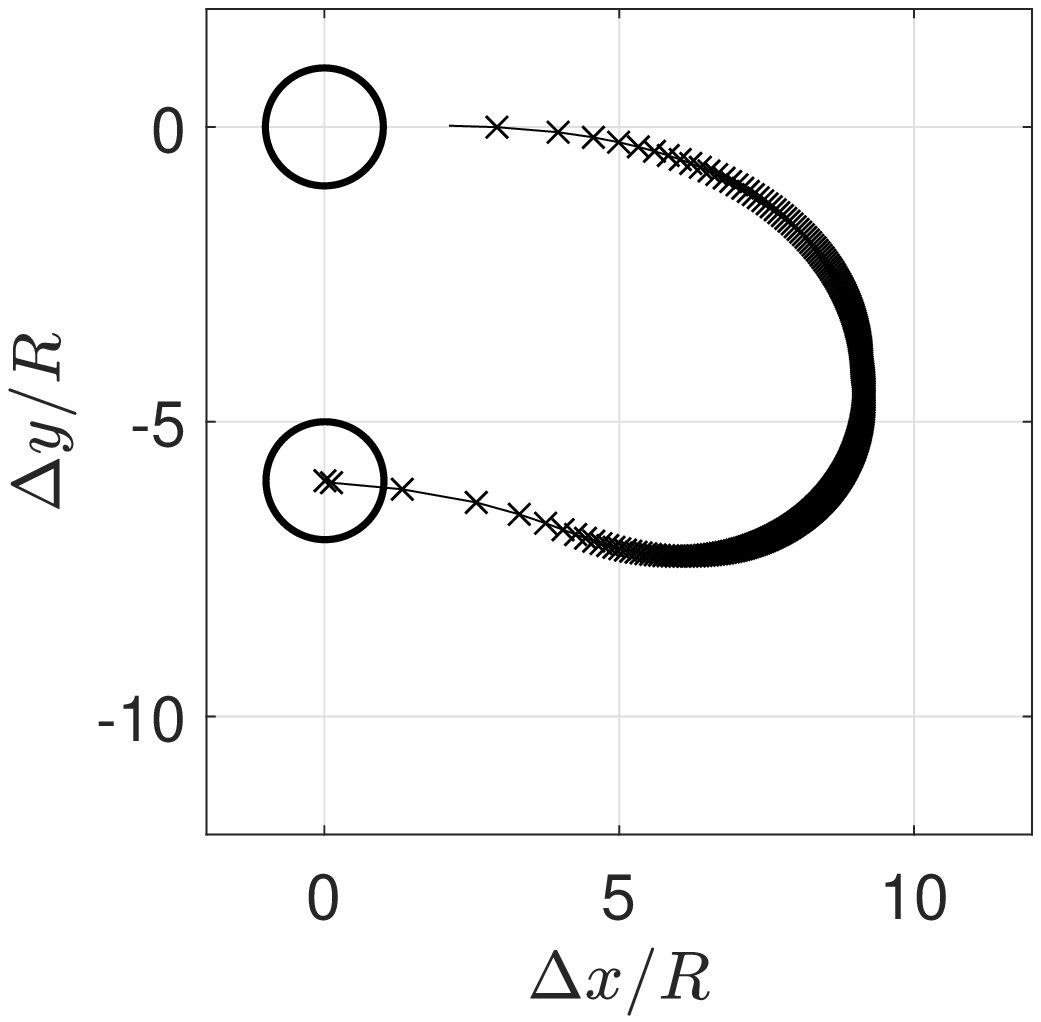}}
  \subfloat[]{\includegraphics[width=55mm,trim=25 0 45 0, clip]{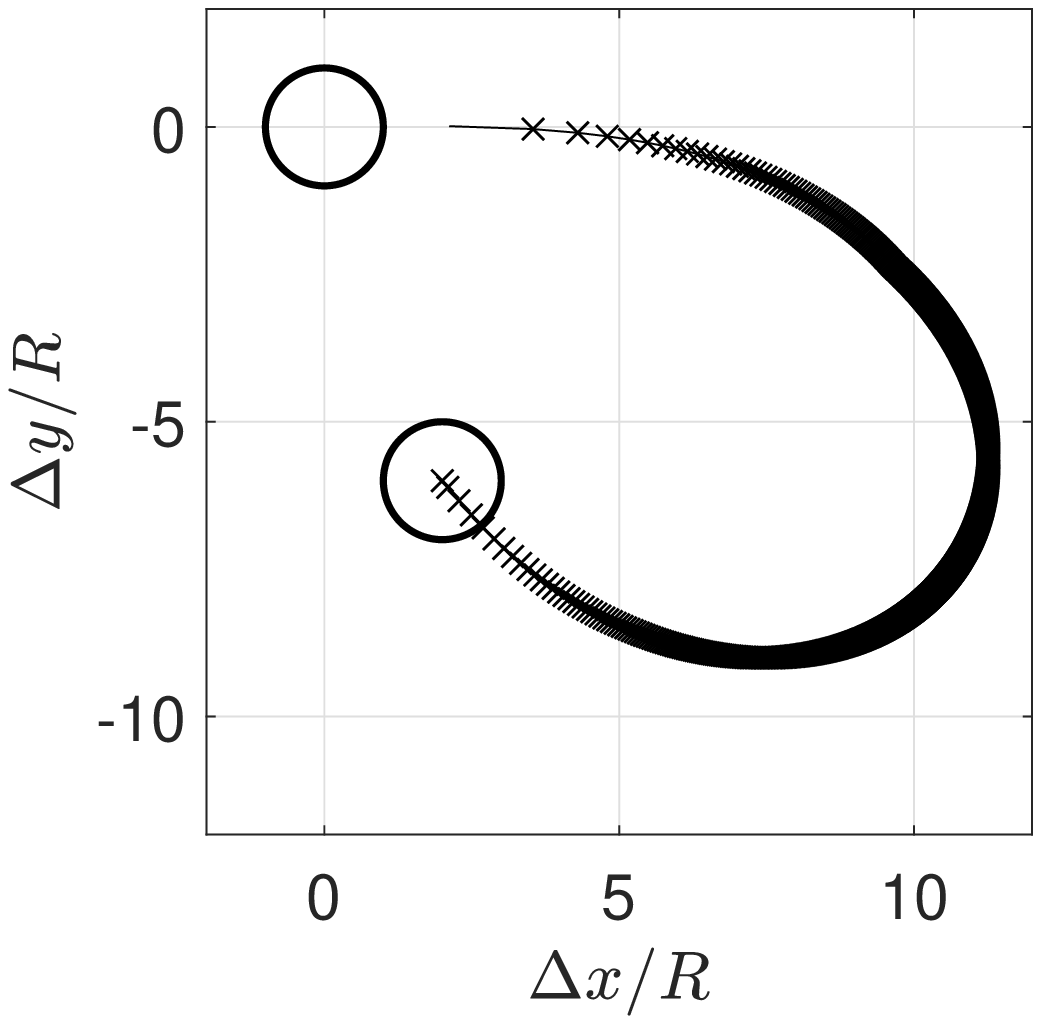}}
  \caption{Trajectories of the trailing bubble relative to the leading bubble with initial configurations of $(\Delta x_0,\Delta y_0)$= (a)$(\varepsilon,6R)$ and (b)$(2R,-6R)$, with viscous interaction. Markers are placed every 5.0 ms.}
\label{fig:traj2}
\end{figure}
To assess the effects of the viscous interaction on the velocity of the trailing bubble, figure \ref{fig:traj2}a and b respectively show the trajectories with its initial positions located at $(\Delta x_0, \Delta y_0)=(\varepsilon,-6R)$ and $(2R,-6R)$, with markers placed every $5.0\times10^{-3}$ s.
In the former initial condition, the trailing bubble is initially immersed in the wake region, while in the latter condition it is out of the wake region.
In figure \ref{fig:traj2}a, the markers indicate that during the initial horizontal motion, the trailing bubble attains its maximum velocity of $O(0.1)$ ms$^{-1}$, relative to the leading bubble.
The motion becomes slower on the subsequent curve-line part of the trajectory, and then it is re-accelerated as the inter-bubble distance becomes small.
The initial velocity in figure \ref{fig:traj2}b is smaller by an order of magnitude than that in figure \ref{fig:traj2}a up to $\Delta x_0\approx 4R$, while the velocities on the later part of the trajectory is similar. The difference in the initial velocity indicates the large amplitude of the wake-induced lift force compared to the force due to potential interaction.
In experiments, it is unlikely that pairs of bubbles are observed in the in-line configuration with the aforementioned equilibrium distance since small disturbances are always present which can immediately trigger this symmetry breaking. This implication agrees with the instability of the in-line configuration observed in the experiments by \citet{Katz96,Sanada05,Kusuno19}.

\begin{figure}
\centering
  \subfloat[]{\includegraphics[width=45mm,trim=80 0 110 0, clip]{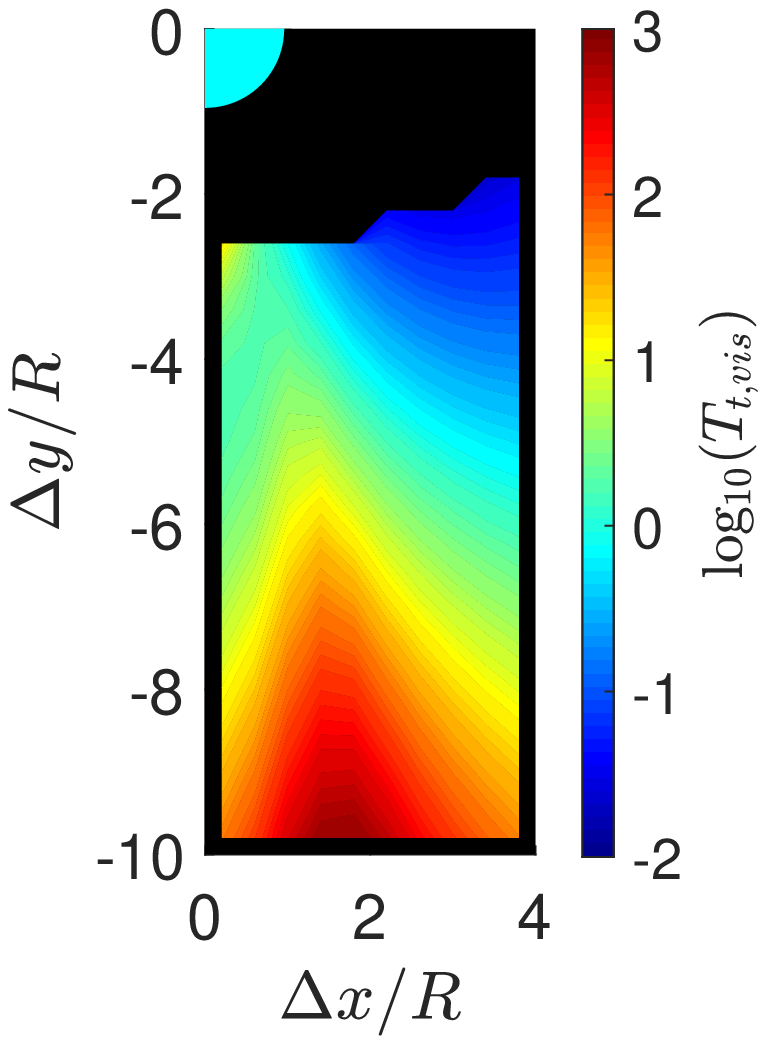}}
  \subfloat[]{\includegraphics[width=45mm,trim=80 0 110 0, clip]{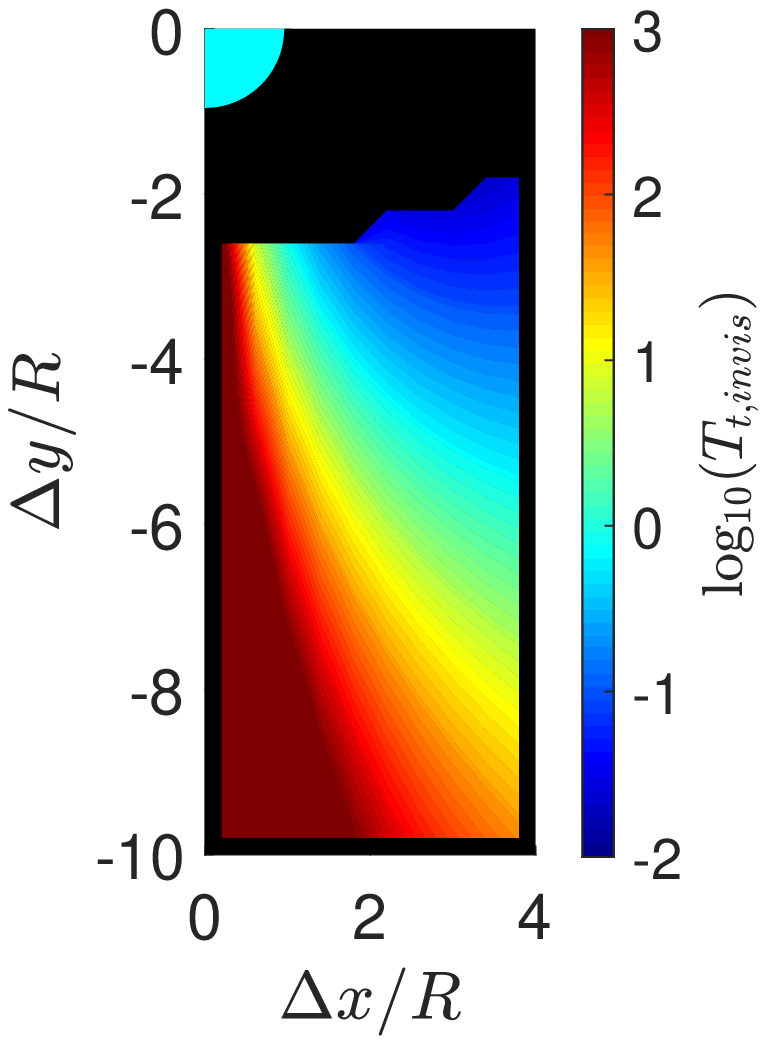}}
  \subfloat[]{\includegraphics[width=45mm,trim=80 0 110 0, clip]{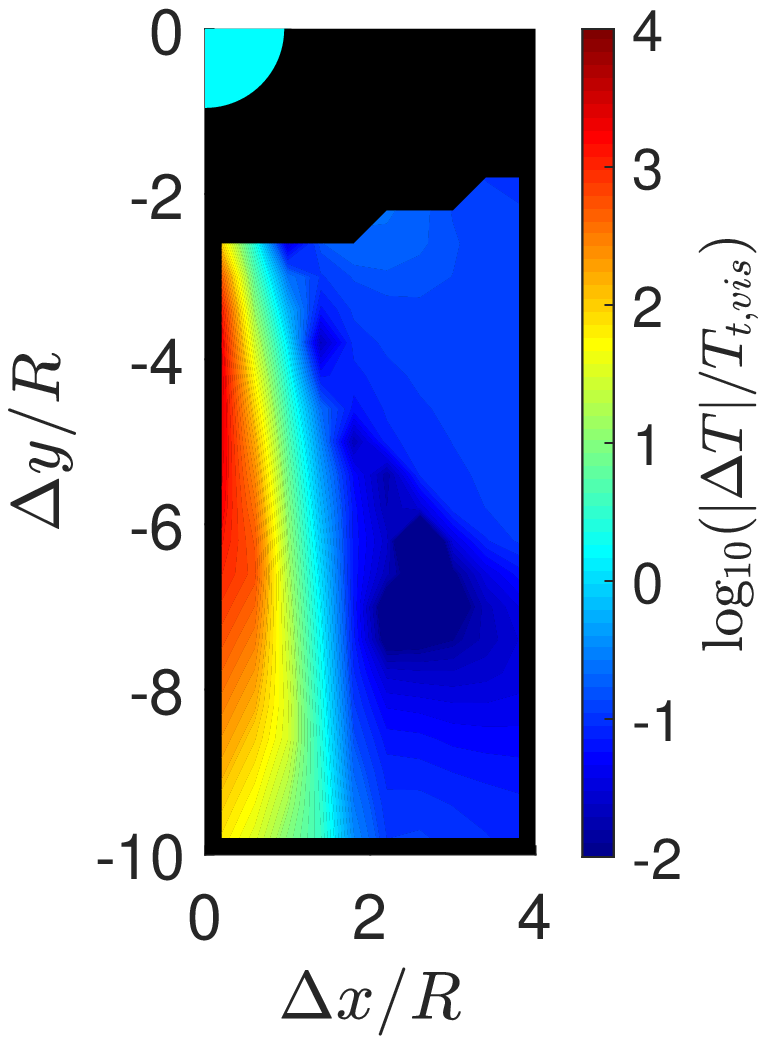}}
  \caption{Contours of the time required for the trailing bubble to catch up with the leading bubble released in the domain without agitation. Results are shown simulated (a) with viscous interaction, $T_{t,vis}$, and (b) without viscous interaction, $T_{t,invis}$. Their normalized difference, $|T_{t,invis}-T_{t,vis}|/T_{t,vis}$, is shown in (c).}
\label{fig:cont_wake}
\end{figure}
To assess the effect of viscous interaction on the total travel time of the trailing bubble until clustering with the leading bubble, in figure \ref{fig:cont_wake}a and b we respectively show contours of the travel time simulated with and without viscous interaction, when released at various relative coordinates. The agitation was not modeled in these simulations.
For both cases, the travel time tends to increase with both the pairwise angle and the inter-bubble distance.
With viscous interaction, the travel time is greater for initial configurations slightly out of in-line than that in perfectly in-line configurations. The maximum travel time in the domain is $O(10^3)$ s around at $(\Delta x,\Delta y)=(2R,-10R)$. Without viscous interaction, the travel time monotonically increases with the pairwise angle for any given inter-bubble distance.
Figure \ref{fig:cont_wake}c shows a contour of the difference between the cases with and without viscous interaction, normalized by the travel time in the case with viscous interaction.
The region with the difference greater than unity is confined in the wake region, $\Delta x:\Delta x<2R$.
On the $\Delta y$-axis, with decreasing $\Delta y$, the difference increases toward $\Delta y_0 \approx -6R$ to take the peak value of $O(10^3)$, and then decays.
In the far-field, the difference is expected to converge to zero as the interaction force becomes infinitesimally small.
Overall, results in figure \ref{fig:cont_wake} highlight the significant acceleration of the side-by-side clustering of in-line pairs by the wake-induced lift force acting on the trailing bubble.

\subsection{Effect of turbulent agitation}
We analyze the effects of the modeled agitation on the pairwise dynamics.
We first address the fluctuation in the travel time of the trailing bubble during the wake-induced translation.
\begin{figure}
\subfloat[]{\includegraphics[width=67mm,trim=0 0 0 0, clip]{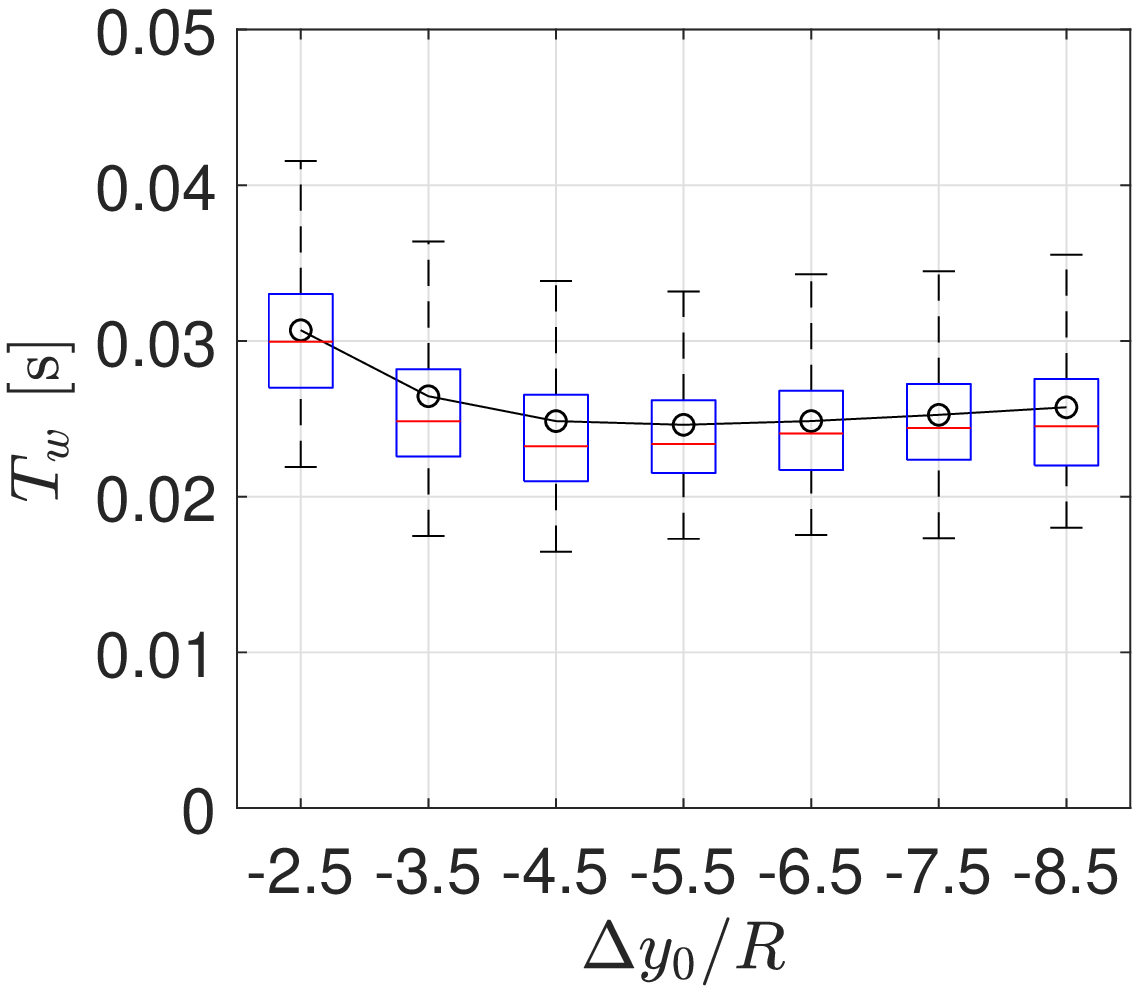}}
\subfloat[]{\includegraphics[width=67mm,trim=0 0 0 0, clip]{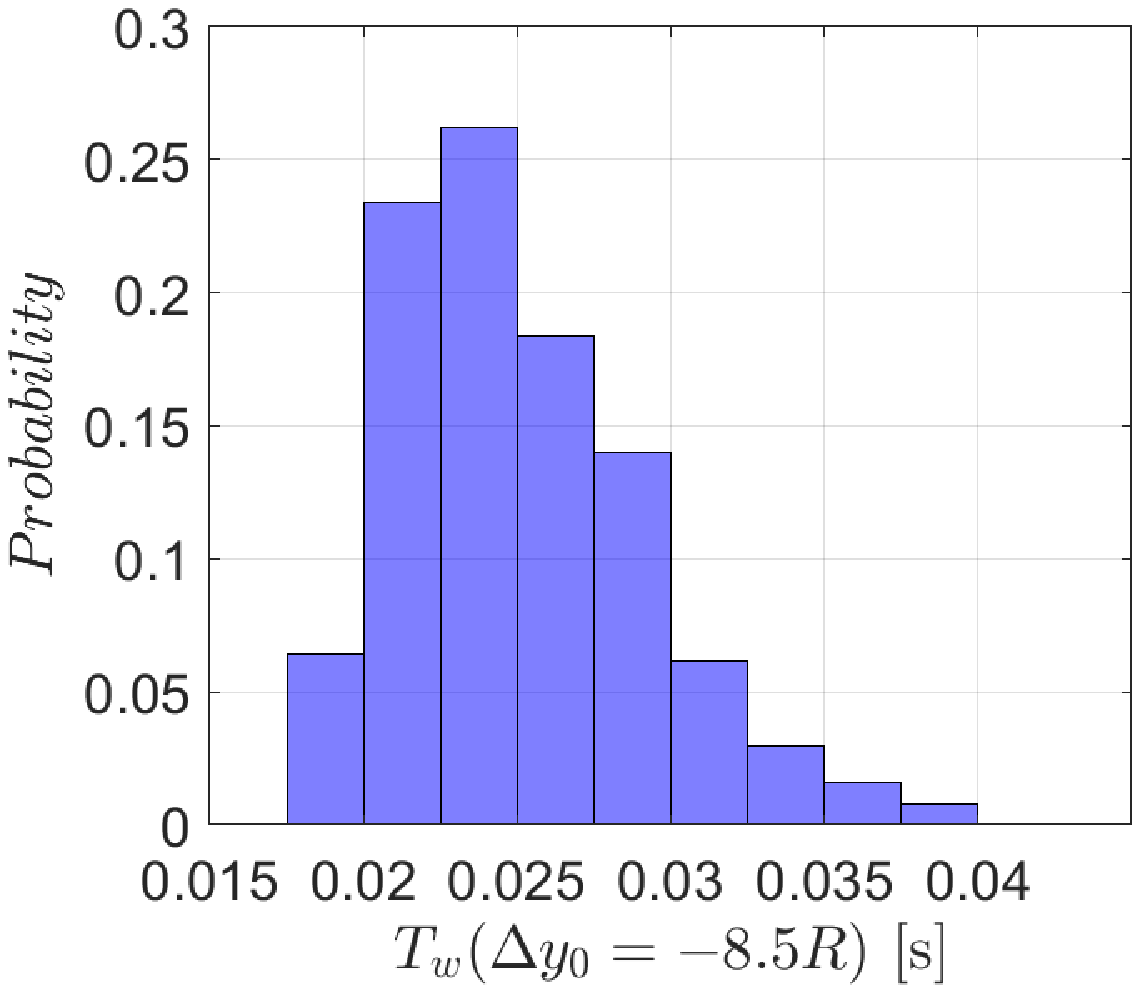}}
  \caption{(a) Box plots of the time required for the trailing bubble to reach $\Delta x=2R$, when released at $\Delta x_0=\varepsilon$ and $\Delta y_0:\Delta y_0 \in[-8.5R, -2.5R]$. Results are obtained with modeling agitation. ($-\circ -$) denotes results without agitation. (b) Probability distribution of $T_s$ for $\Delta y_0=-8.5R$.}
\label{fig:timeb}
\end{figure}
Figure \ref{fig:timeb}a shows the box plot of the travel time required for the trailing bubble to reach $\Delta x=2R$, when it is released at $(\Delta x, \Delta y)=(\varepsilon,\Delta y_0)$, where $\Delta y_0 \in[-8.5R, -2.5R]$.
At all $\Delta y_0$, the travel time is on the order of $O(10^{-2})$ s.
This timescale is as small as the minimum timescale of the fluctuation, $1/f_c$, and the bubble travels on a ballistic path during the simulation.
The median travel time is slightly smaller than that of the case without agitation at all $\Delta y_0$. The median of the travel time slightly decreases with decreasing $\Delta y_0$ up to $\Delta y_0=-4.5 R$ and then becomes nearly constant at smaller $\Delta y_0$.
The fluctuation of the travel time is spread within 80\% of the mean travel time at all $\Delta y_0$.
Figure \ref{fig:timeb}b shows the probability distribution of the travel time for the case with $\Delta y_0=-2.5R$. The distribution presents a skewness of 0.85.
This positive weak skewness can be explained by the growth of the potential drag $C_{D\xi,int}$ with increasing the horizontal velocity of the trailing bubble. Overall, results shown in figure \ref{fig:timeb} indicate that the timescale of the wake-induced translation is small and not influenced by the agitation, regardless of the inter-bubble distance considered.

Next, we address the fluctuation during the trailing bubble traveling on the curve-line paths driven by potential interaction, for oblique pairs out of the influence of the wake.
\begin{figure}
\subfloat[]{\includegraphics[width=67mm,trim=0 0 0 0, clip]{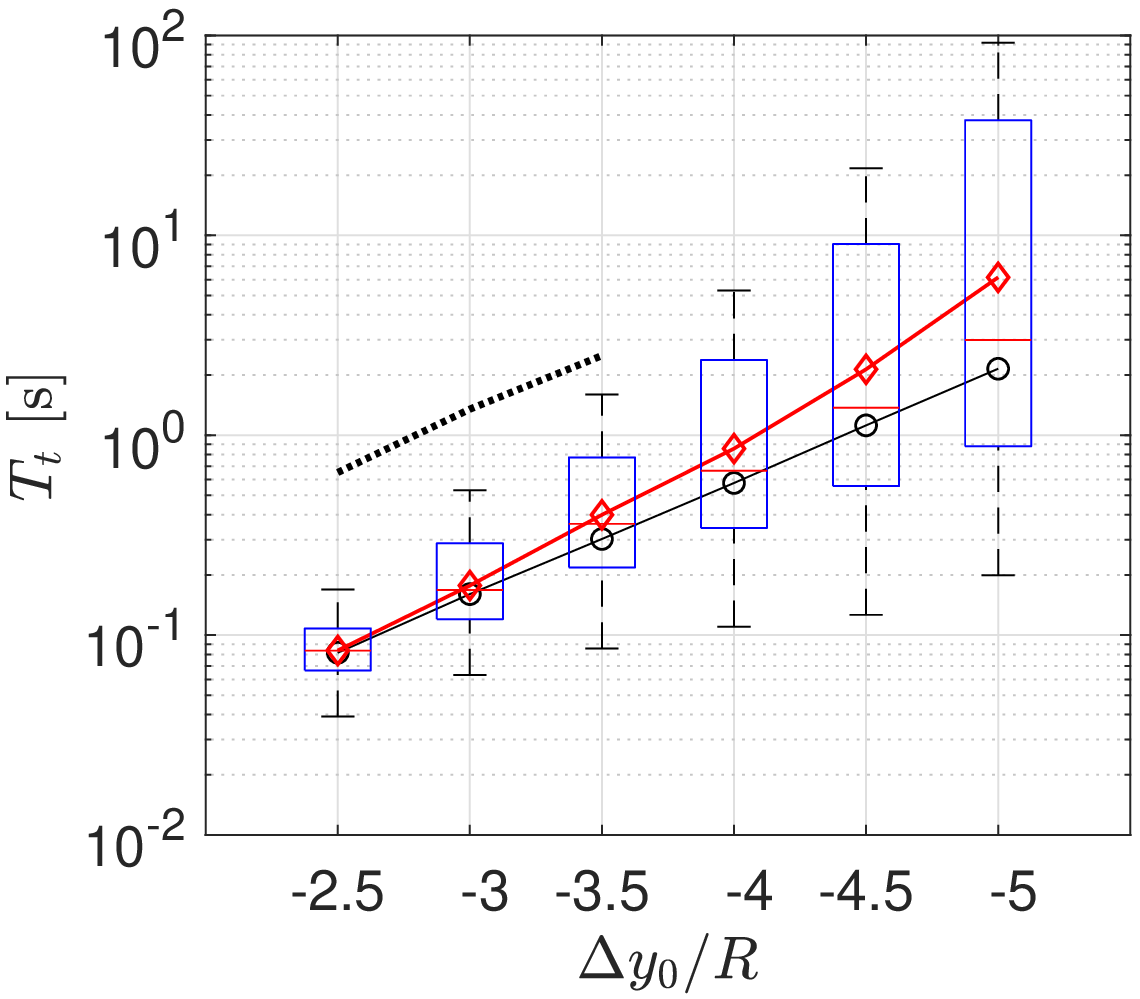}}
\subfloat[]{\includegraphics[width=67mm,trim=0 0 0 0, clip]{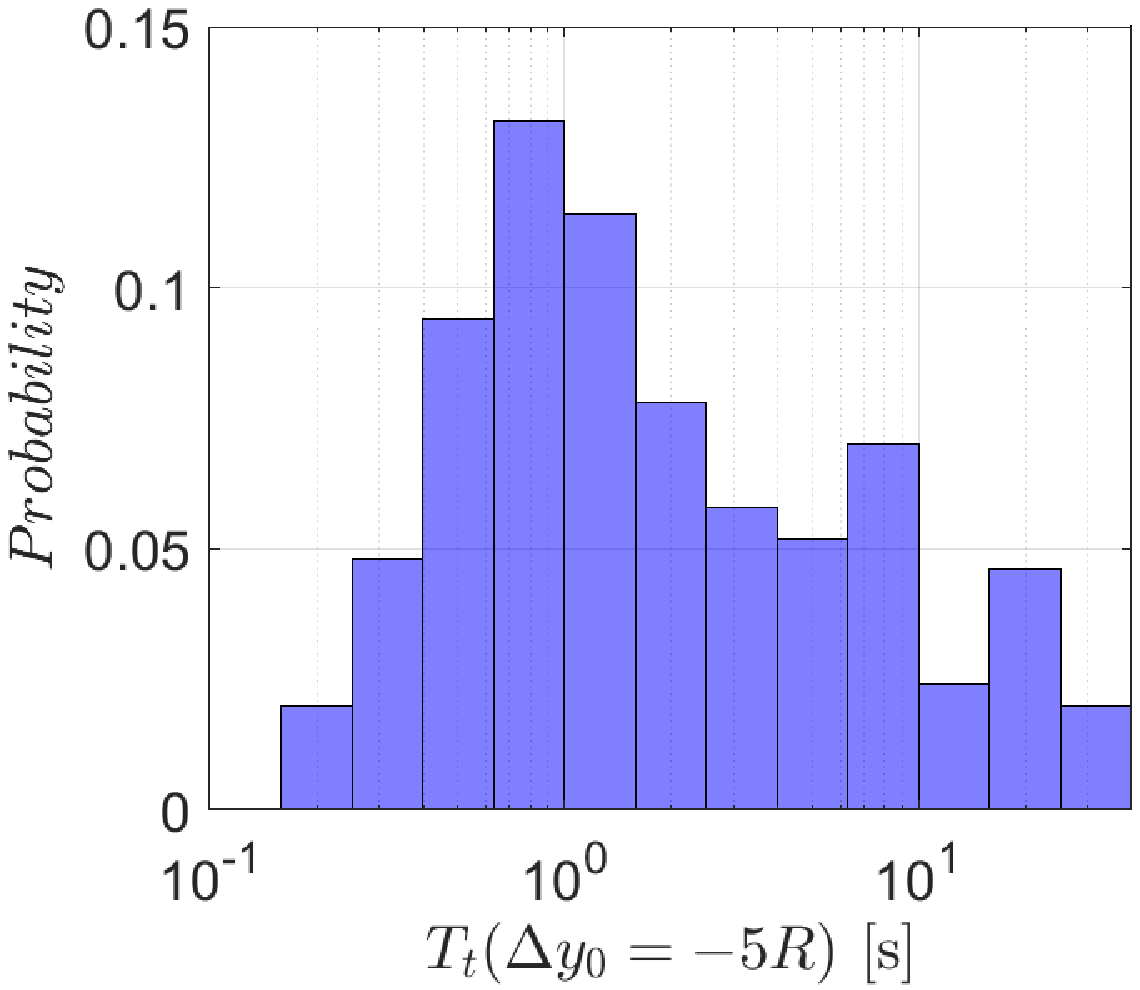}}
  \caption{(a) Box plots of the time required for the trailing bubble to catch up with the leading bubble, when it is released at the coordinate of $\Delta x_0=2R$ and $\Delta y_0:\Delta y_0 \in[2.5R, 6.5S]$.
  ($-\circ -$) denotes results without agitation. ($-\diamond-$) denotes the box mean. ($--$) denotes $T_t\sim \Delta y_0^4$.
  (b) Probability distribution of $T_s$ for $\Delta y_0=-5R$, up to the third quantile.}
\label{fig:timec}
\end{figure}
Figure \ref{fig:timec}a shows the box plot of the time required for the trailing bubble to catch up with the leading bubble, when the bubbles are released at $\Delta x_0=2R$, with various values of $\Delta y_0$.
In these conditions, the trailing bubble is subjected almost only to the potential-induced attraction and the agitation.
The travel time without agitation scales like $|\Delta y_0|^4$. This scaling agrees with the theoretical prediction of equation (\ref{eqn:final}), based on the leading order of potential interaction, $S^{-4}$.
The magnitude of the fluctuation (box height) significantly increases with decreasing $\Delta y_0$. 
The median values of the travel time in the case with agitation do not significantly deviate from that without agitation.
The mean travel time, on the other hand, positively deviates from the travel time without agitation with a greater magnitude compared to the median. The difference between the median and the mean also increases with decreasing $\Delta y_0$. The increase in the deviation between the median and the mean suggests the growth of the skewness of the fluctuation.
Figure \ref{fig:timec}b shows the probability distribution of the travel time at $\Delta y_0=-5R$, up to the third quantile.
As predicted, the distribution is highly non-Gaussian; it spreads over greater than two decades and its profile is skewed even on the semi-logarithmic axis.

\begin{figure}
\centering
  \subfloat[]{\includegraphics[width=45mm,trim=40 0 45 0, clip]{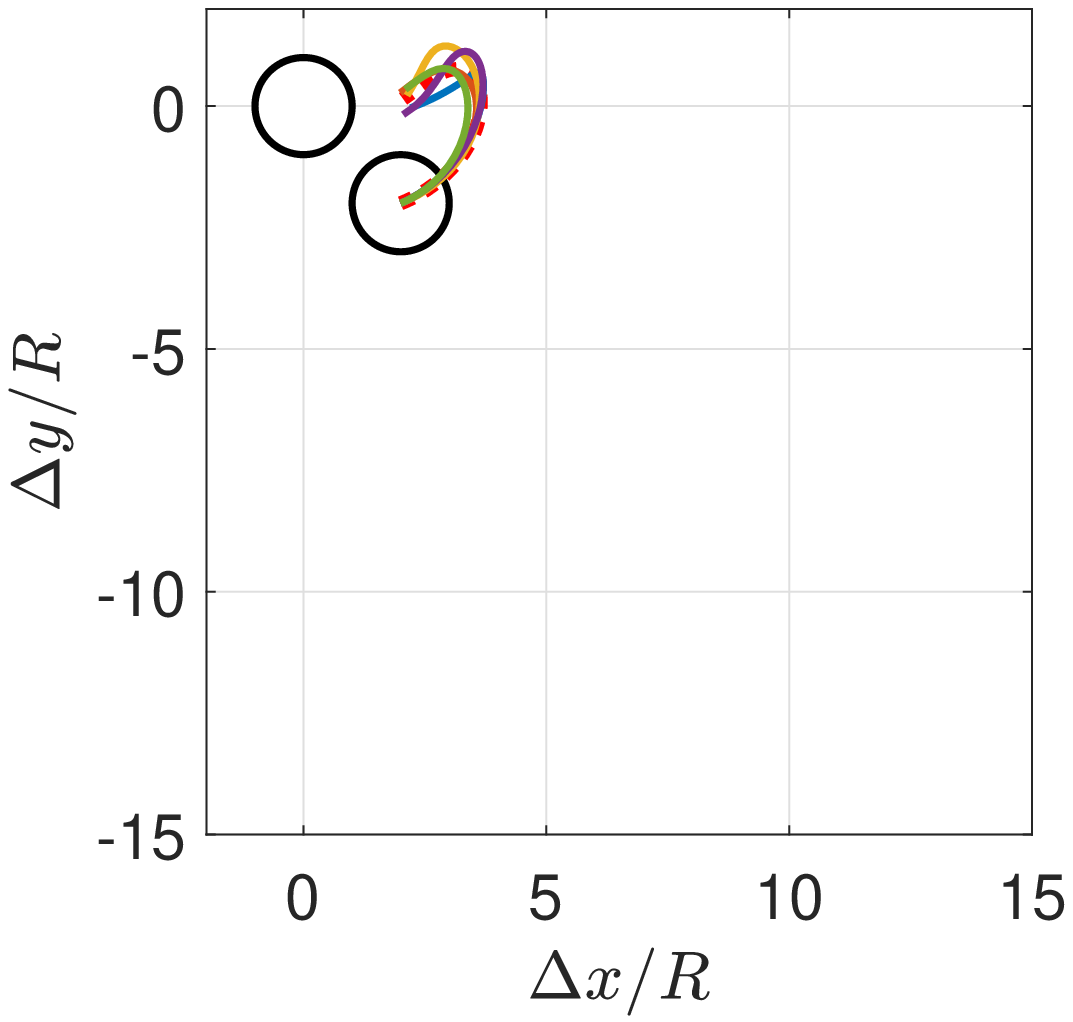}}
  \subfloat[]{\includegraphics[width=45mm,trim=40 0 45 0, clip]{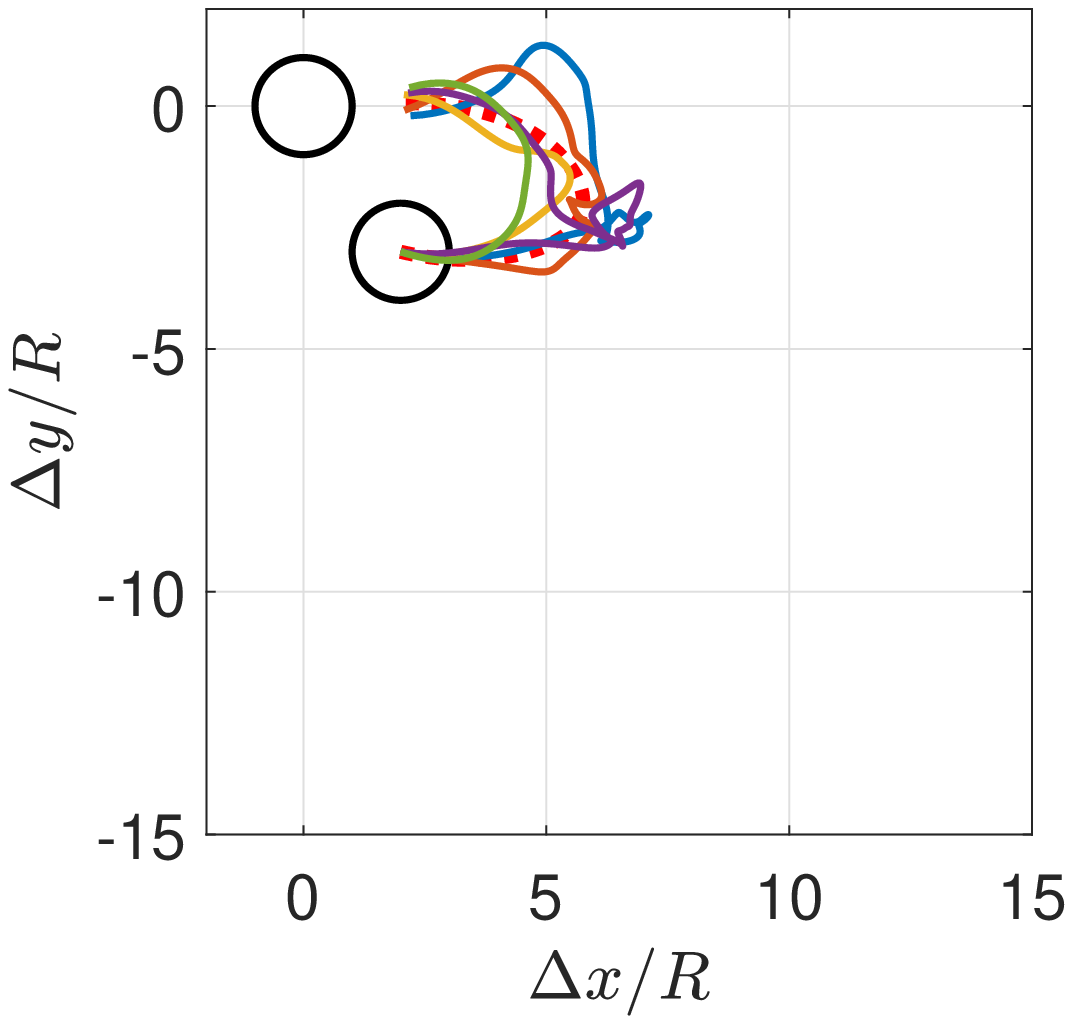}}
  \subfloat[]{\includegraphics[width=45mm,trim=40 0 45 0, clip]{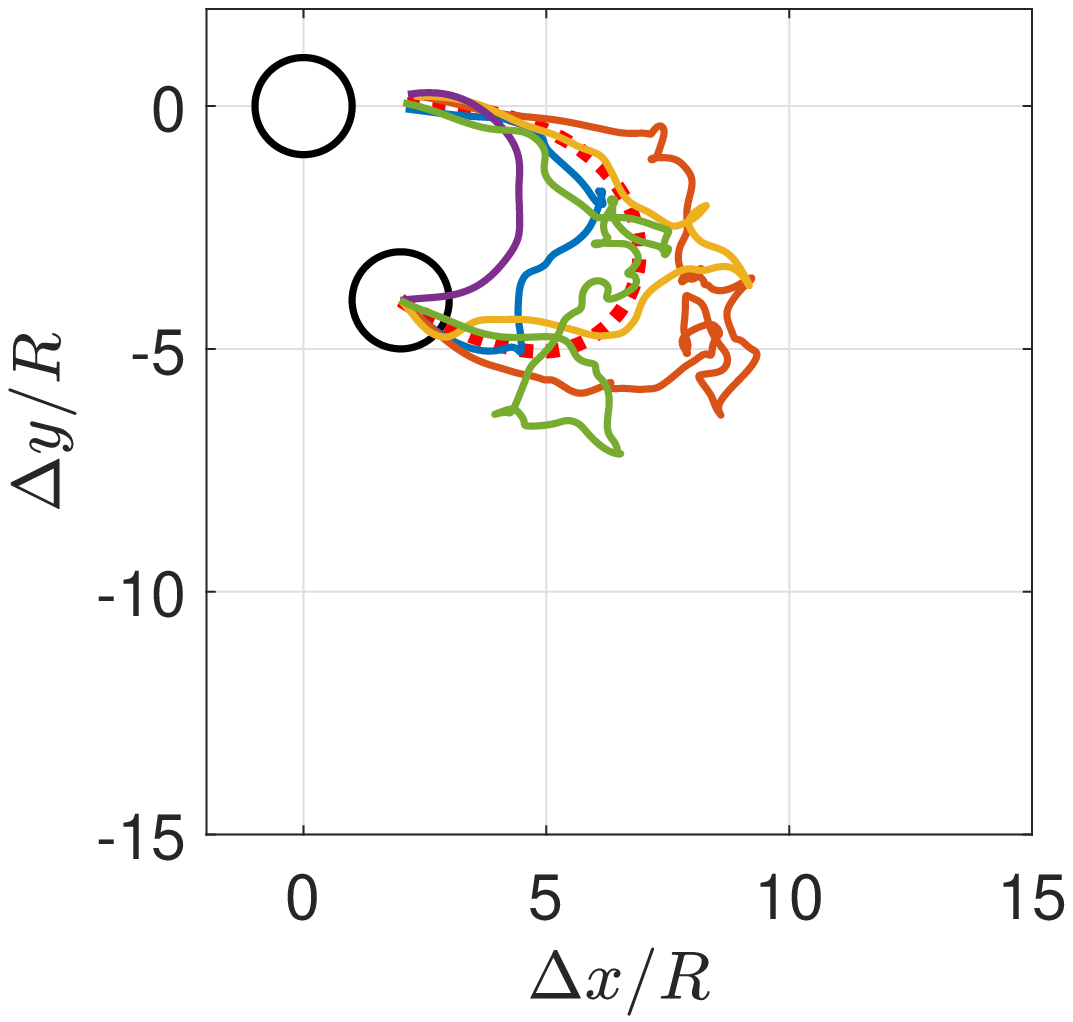}}
  \caption{Representative trajectories of the trailing bubble relative to the leading bubble, with viscose interaction and agitation. The initial positions of the trailing bubble are $(\Delta x_0, \Delta y_0)=$ (a)$(2R, -2.0R)$, (b)$(2R, -3.0R)$, and (c)$(2R, -4.0R)$.
  }
\label{fig:trajb}
\end{figure}
To gain further insights on the fluctuation in the travel time, figure \ref{fig:trajb} shows representative trajectories of the trailing bubble relative to the leading bubble simulated with viscous interaction and agitation, with various initial configurations.
While fluctuated, the trajectories are overall similar to the cases without agitation.
The magnitude of the fluctuations, however, apparently become greater for pairs with a larger initial inter-bubble distance, suggesting that the fluctuation in the travel time results from that of the trajectories. Compared to the pairwise bubbles without agitation, these results also suggest that the near in-line pairs in turbulence can reach side-by-side within a timescale an order or magnitude smaller and larger with non-negligible probabilities.
The enhancement of the fluctuation in both the travel time and the trajectory against the increase in the inter-bubble distance can be explained by the growth of the relative magnitude of the agitation compared to the potential flow attraction; the amplitude of the agitation is homogeneous, while the amplitude of potential interaction rapidly decays with increasing the inter-bubble distance.
The significant acceleration of the side-by-side clustering by the additive fluctuation may be counter intuitive, compared to the deceleration. Our interpretation is that the acceleration is due to the highly path-dependent nature of the travel time. With a finite probability, the agitation can drive the trailing bubble closer to the leading bubble. In that case, the trailing bubble moves to an inner curve-line trajectory with a smaller radius on which the bubble travels faster. For pairs at a shallow angle, fluctuations can also drive the trailing bubble upper relative to the leading bubble, and this motion can trigger side-by-side potential attraction earlier than the case without agitation, resulting in faster clustering.

The timescales of the wake-induced symmetry breaking and the potential-driven side-by-side clustering identified in this section, and their comparisons with the travel time of bubbles in the experimental imaging, $O(1)$ s, explain the preferred configurations captured in the images.
The timescale of the initial symmetry breaking is $O(10^{-2})$ s, regardless of the agitation.
The in-line configuration is therefore likely to be broken before reaching the experimental test section. 
Pairs initially at $S<5$ reach side-by-side and those at $S>5$ remain in the near in-line configurations, within the timescale of $O(1)$ s.
Therefore, assuming that the distribution of the pairwise angle is initially uniform, the probability of observing the near in-line pairs is increased and that of the perfect-inline pairs is decreased, for pairs at $S>5$ in the test section.
On the other hand, pairs initially at $S<5$ take side-by-side configurations before they reach the test section.

\section{Connections with the cluster formation}\label{sec:types_paper}
\subsection{Onset of dilute cluster formation in the experiment}
To further discuss implications of the pairwise dynamics for the mechanism of the formation of larger bubble clusters, we capture the onset of clustering of multiple bubbles in the channel using the same experimental setup.
The bulk flow condition and the surfactant concentration follow \S\ref{sec:exp}.
This time the air is continuously supplied to realize an average void fraction of 0.1\%.
At this void fraction, the average inter-bubble distance becomes approximately $6R$, assuming that all bubbles are trapped by the wall.
To capture the evolution of the spatial distributions of bubbles, we captured instantaneous images of bubbles at three different test sections located at various heights from the bubble inlet: $h=0.1$, $0.3$, and $0.5$ m. The time required for the bubbles to reach the center of each test sections from the inlet is $t=0.5$, 1.0, and 2.5 s, respectively.

\begin{figure}
 \begin{center}
\subfloat[]{
 \includegraphics[width=42mm]{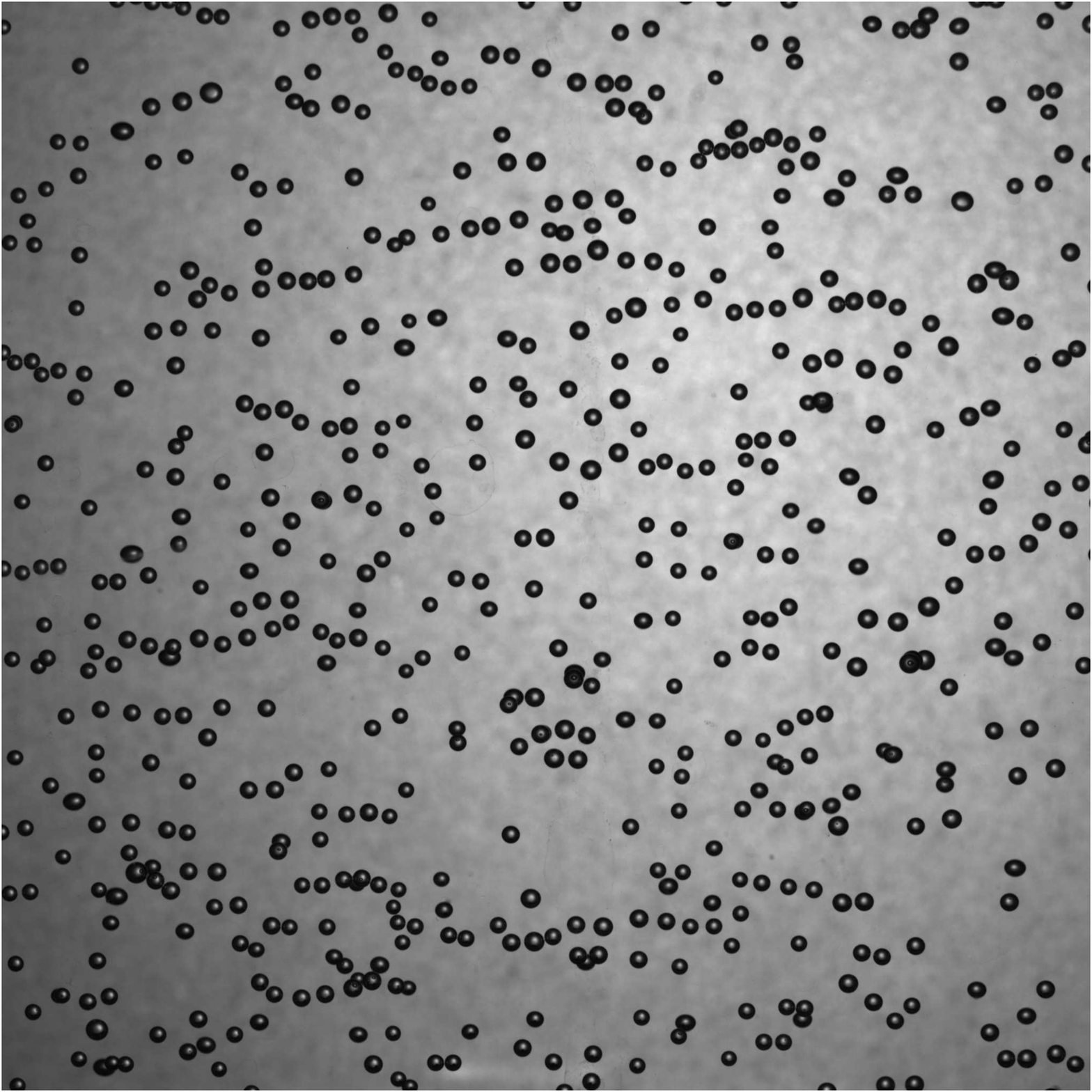}
 \label{right1}
 } 
\subfloat[]{
 \includegraphics[width=42mm]{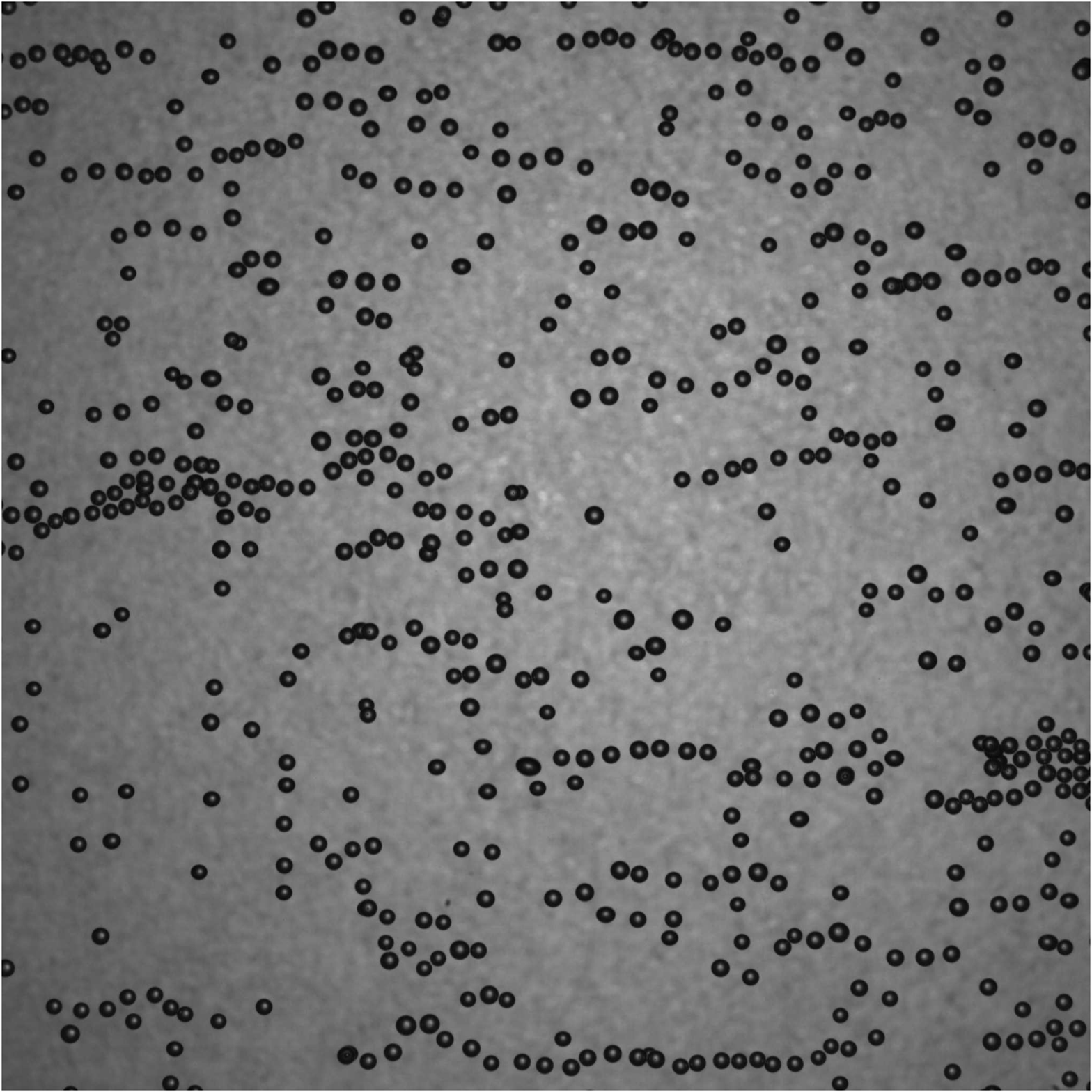}
 \label{right2}
 } 
\subfloat[]{
 \includegraphics[width=42mm]{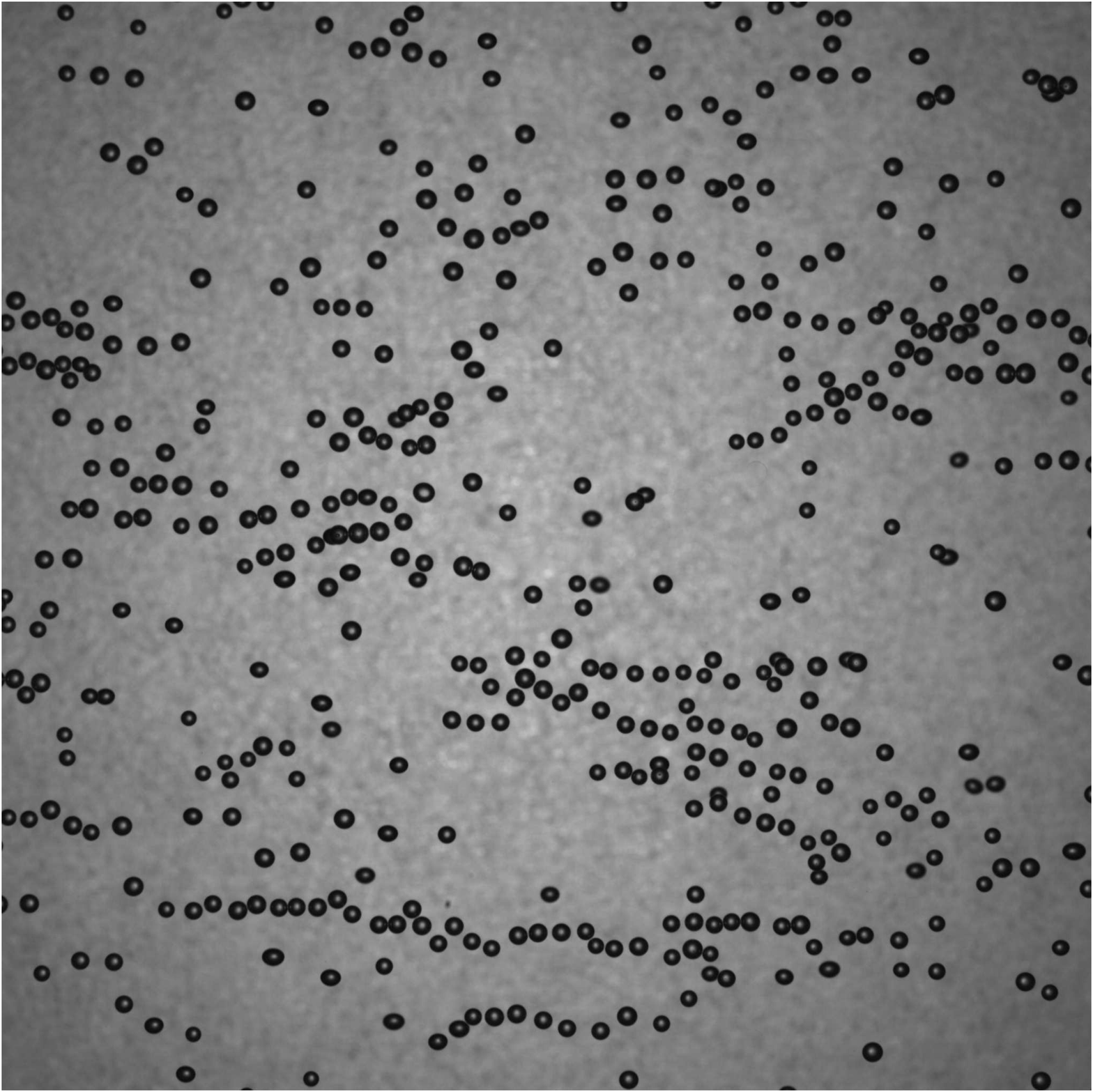}
 \label{right3}
 } \\
 
 \subfloat[]{
 \includegraphics[width=42mm,trim=0 0 20 0,clip]{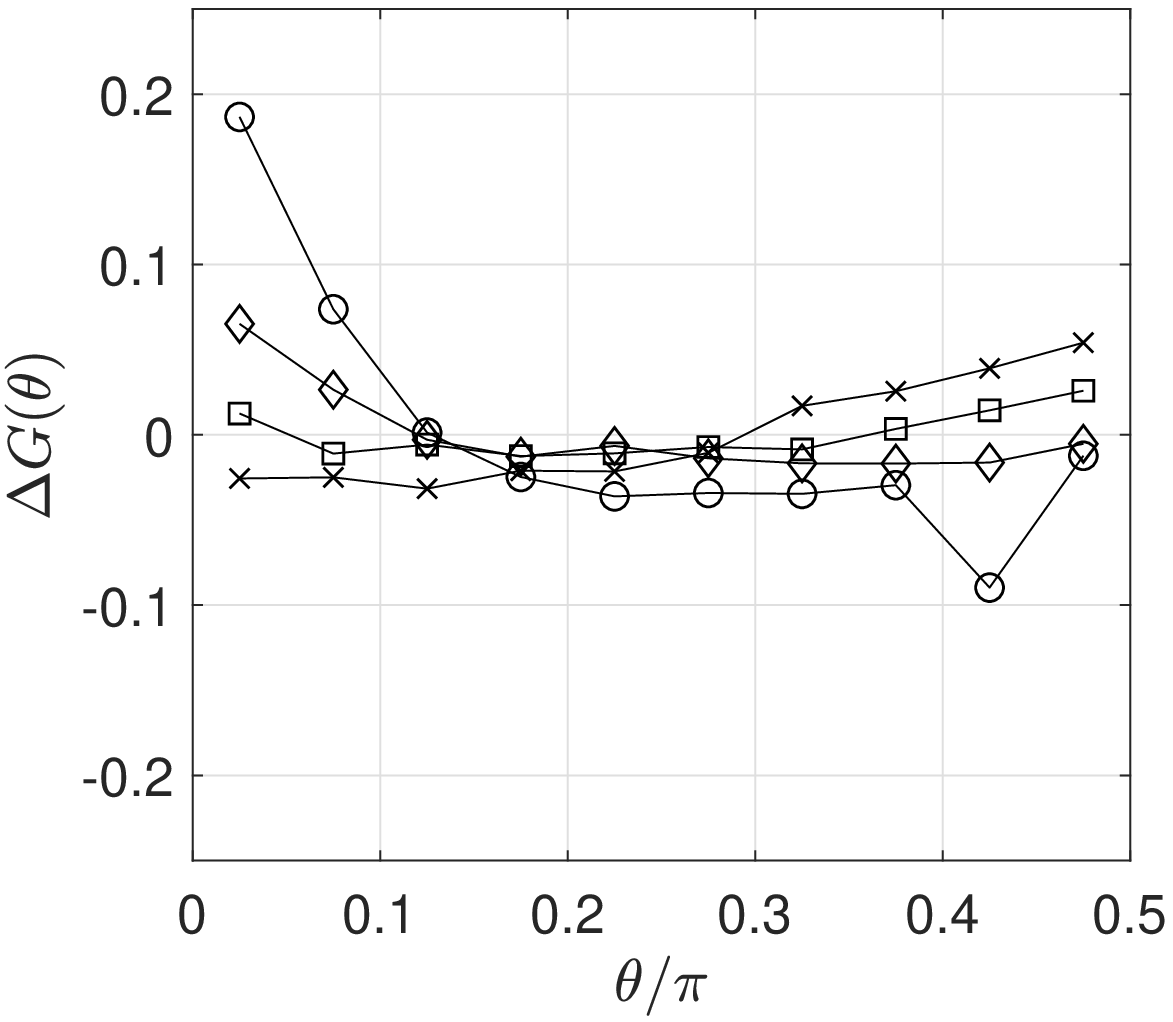}
 \label{left1}
 } 
\subfloat[]{
 \includegraphics[width=42mm,trim=0 0 20 0,clip]{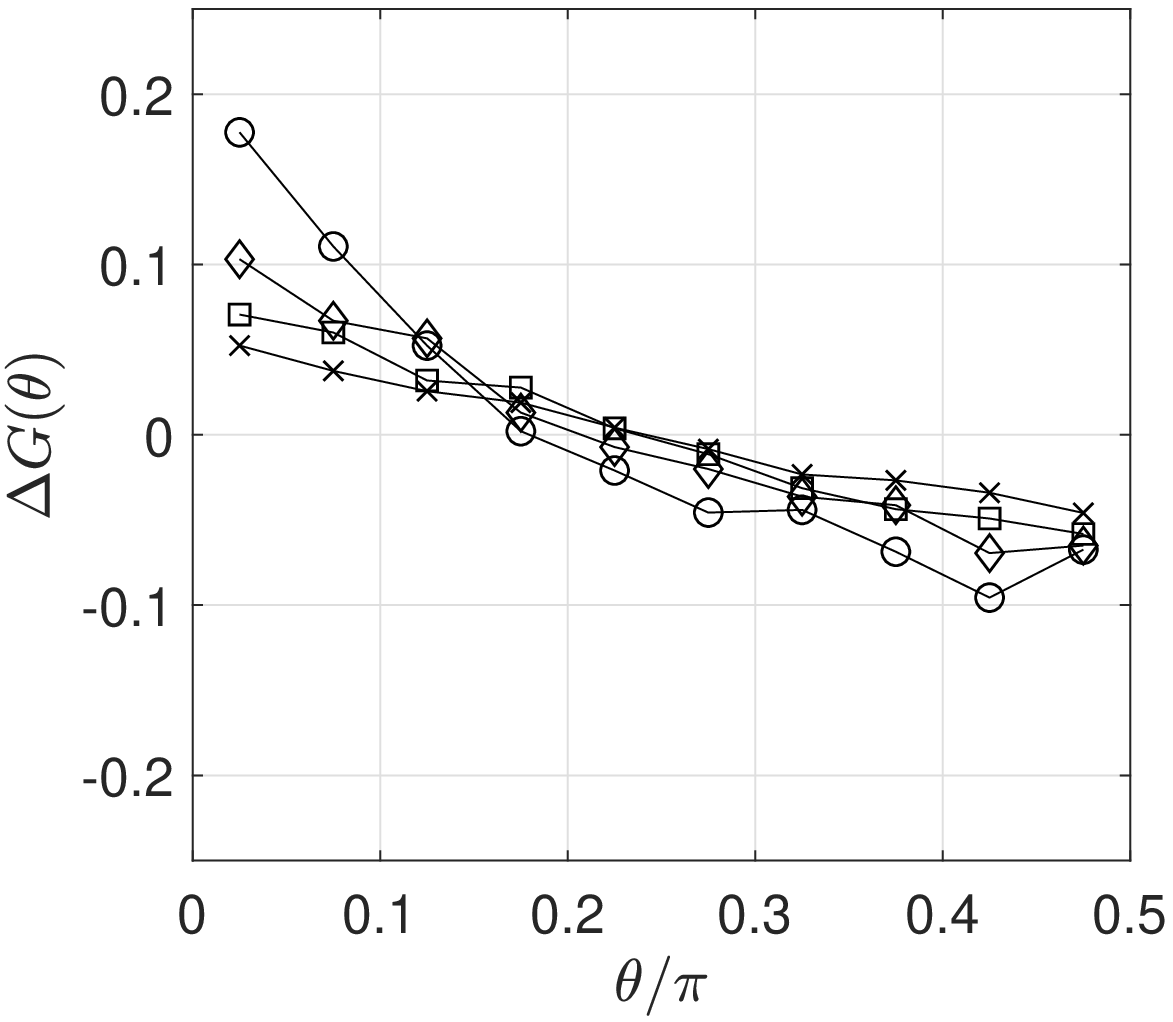}
 \label{left2}
 }
\subfloat[]{
 \includegraphics[width=42mm,trim=0 0 20 0,clip]{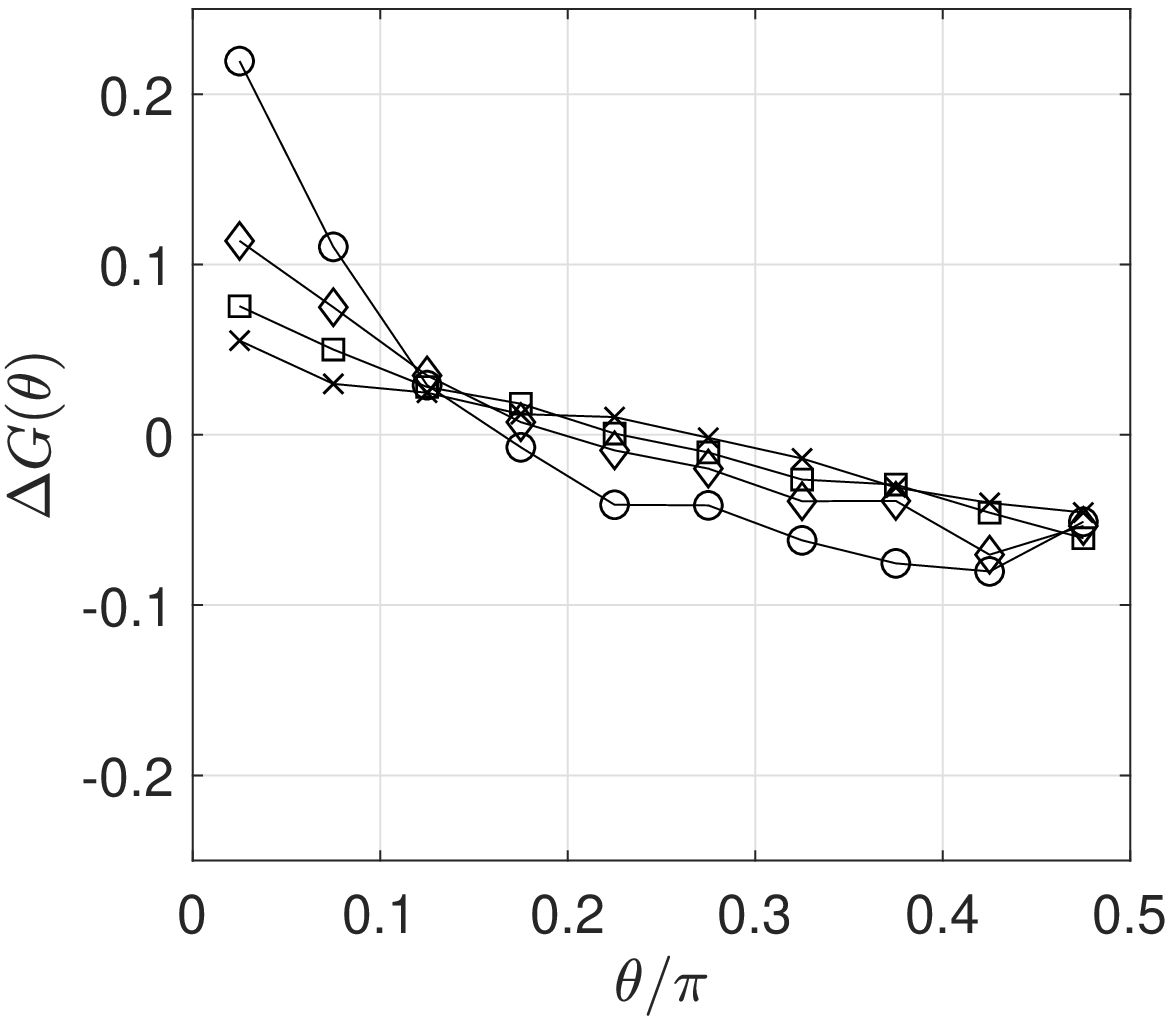}
 \label{left3}
 }
 \caption{(a-c) Images of the bubbles captured in the test sections located at different heights. (d-f) Angular pair distribution function of the bubbles obtained from the sequence of images the sections corresponding to (a-c). Symbols: $\circ$, $S=5.0$; $\diamond$, $S=7.5$; $\square$, $S=10.0$; $\times$, $S=12.5R$.}
 \label{fig:cluster}
 \end{center}
\end{figure}
Figure \ref{fig:cluster}a-c show the images of bubbles obtained at the test sections.
To quantify the state of clusters, we compute the pair probability distribution function (PPDF):
\begin{align}
\Gamma(S, \theta)
&=
\frac{\Omega}{N_b(N_b-1)}\sum_{i=1}^{N_b}\sum_{j=1, j\neq i}^{N_b}\delta(S-|S_{ij}|)\delta(\theta-|\theta_{ij}|),
\end{align}
where $N_b$ is the total number of bubbles in each image and $\Omega$ is normalization constant.
We approximate the delta functions following the approach used to obtain C-PPDF.
Notice that the physical interpretation of PPDF is distinct from C-PPDF, since to obtain PPDF all bubbles are sampled from each image without conditioning.
To obtain statistical mean, we average $\Gamma$ using over $10^3$ images at each test section without overlap.
For convenience, from PPDF we compute the residual of the angular pair distribution function (APDF) normalized at each $S$ after subtracting unity:
\begin{align}
\Delta G(\theta)=\frac{\Gamma(S,\theta) \int d\theta}{\int\Gamma(S,\theta) d\theta}-1,
\end{align}
which represents the bias of the probability to observe pairs with a configuration angle $\theta$ at each $S$. If the distribution is uniform at all angles, $\Delta G(\theta)=0$.
Figure \ref{fig:cluster}d-f respectively show $\Delta G(\theta)$ for $S=$5.0, 7.5, 10, and 12.5, which are obtained from the images captured in the corresponding test sections in figure \ref{fig:cluster}a-c.
In figure \ref{fig:cluster}d, $\Delta G$ largely varies with $S$. 
For $S=5.0$, $\Delta G$ takes its peak with values of around 0.2 at $\theta=0$.
$\Delta G$ decays with $\theta$ and becomes negative at $\theta:\theta>0.2\pi$.
For larger $S$, the distribution varies with $h$.
Small peaks are observed at $\theta=0$ for $S=7.5$ and 10. For $S=12.5$, $\Delta G$ presents a weak positive correlation with $\theta$.
These plots indicate that pairs with a short inter-bubble distance ($S<7.5$) tend to become side-by-side, while specific structures are not observed at larger scales.
On the other hand, in both figure \ref{fig:cluster}e and f, the profiles of the plots for various values of $S$ are similar. In all plots in both figures, $\Delta G$ takes its peak values at $\theta=0$ and almost steadily decays with $\theta$. $\Delta G$ takes negative values at $\theta>0.2$ for all plots. The magnitude of the slope decays with increasing $S$.
These profiles of the plots in figure \ref{fig:cluster}e and f indicate that bubbles tend to be aligned horizontally regardless of their mutual distance, at $h=0.3$ and 0.5 m.
Moreover, the resemblance of figure \ref{fig:cluster}e and f, and their clear difference from figure \ref{fig:cluster}d indicate that the evolution of the horizontal clustering reaches its stationary state before bubbles reach $h=0.3$ m after $h=0.1$ m, therefore at $t:0.5<t<1.5$ s.
The angle dependence of $\Delta G$ and its evolution in figure \ref{fig:cluster}d-f agree with the visual orientation of bubbles in figure \ref{fig:cluster}a-c. In figure \ref{fig:cluster}a, many pairs are in side-by-side configurations, while larger clusters are not present. In figure \ref{fig:cluster}b, on the other hand, we clearly observe long, horizontal chains of bubbles. Similar clusters are observed in figure \ref{fig:cluster}c.

\subsection{Probability of pairwise clustering}
The clustering of multiple bubbles involve many-body interactions, and implications from the pairwise interaction are clearly limited. Nevertheless, potential interaction is short-range and the effects of the nearest neighbors are expected to be the most dominant. Therefore, the formation of dilute clusters is plausibly controlled by a finite number of local pairwise interactions, and its timescale can be associated with that of pairwise clustering.

\begin{table}
 \begin{center}
  \begin{tabular}{cccc}
    Set & Interaction  & Agitation \\[3pt]
     A & Potential  & N/A\\
     B & Potential  & Uncorrelated\\
     C & Potential  & Correlated\\ 
     D & Potential+Viscous  & N/A\\
     E & Potential+Viscous  & Uncorrelated\\
     F & Potential+Viscous  & Correlated\\
 \end{tabular}
  \caption{Summary of the model settings for the simulation of the probability of side-by-side clustering.
  }{\label{tab:pm}}
 \end{center}
\end{table}
To assess the possible effects of viscous interaction and the agitation as well as the spatial correlation of the agitation on the onset of cluster formation, we simulate the dynamics of pairwise bubbles with various initial configurations during the clustering timescale identified in the experiment, $O(1)$ s, with various model settings as summarized in table \ref{tab:pm}.
In these sets, we address combinations of the interaction models with and without viscous effects, and agitation models with and without spatial correlation.
In the cases without the agitation (set B,C,E, and F), the initial position of the trailing bubble is defined at the center of $16\times20$ grids that uniformly discretize the domain of $\Delta x\in[0,3.2]$ and $\Delta y\in[-4,0]$.
500 simulations are conducted for the set of $(\Delta x, \Delta y)$ at each grid cell to obtain the probability of side-by-side clustering to occur within time $t$: $p_c(t,\Delta x,\Delta y)$.
In the cases with the agitation, $48\times60$ grids are used for the same domain. A single simulation is conducted at each grid cell.

\begin{figure}
  \subfloat[]{\includegraphics[width=45mm,trim=35 0 65 0, clip]{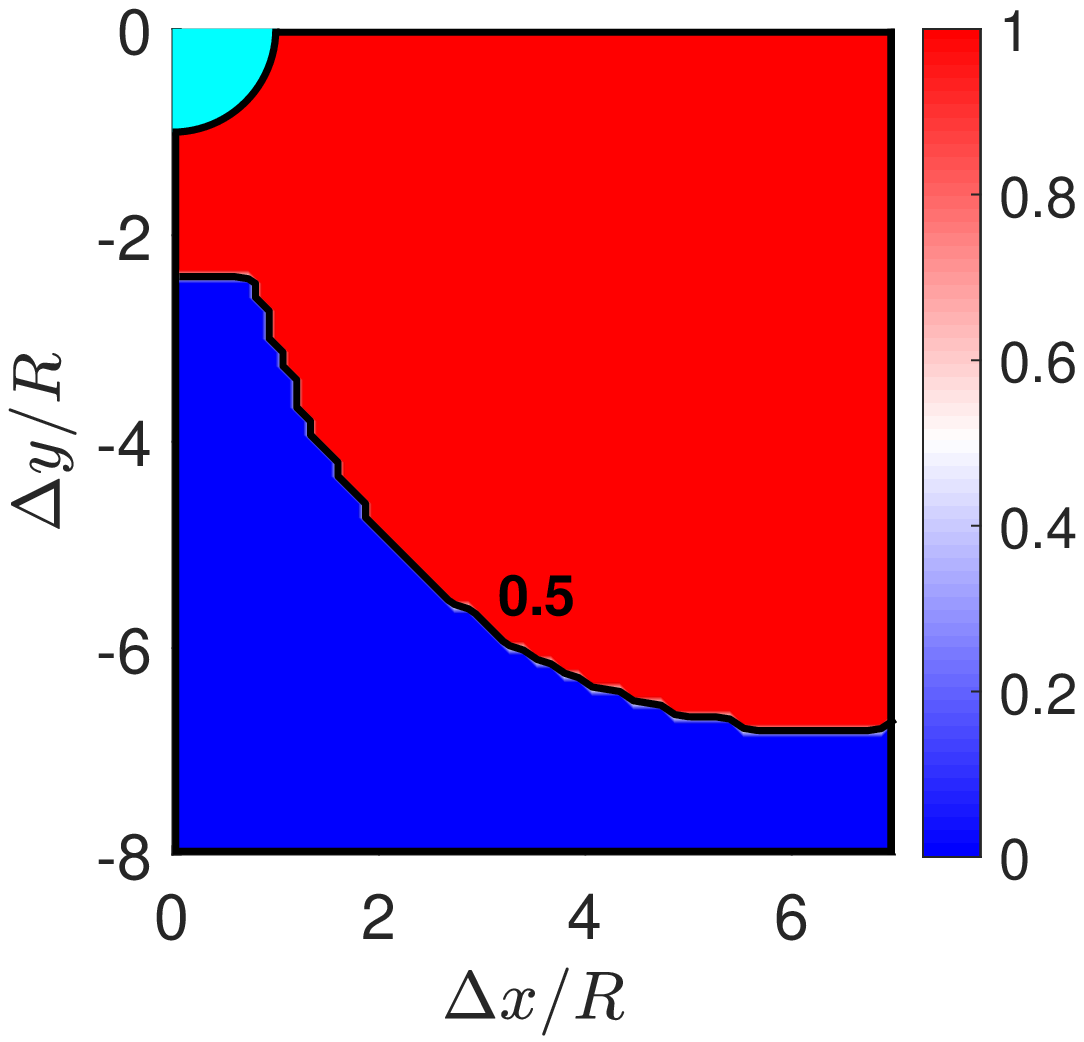}}
  \subfloat[]{\includegraphics[width=45mm,trim=35 0 65 0, clip]{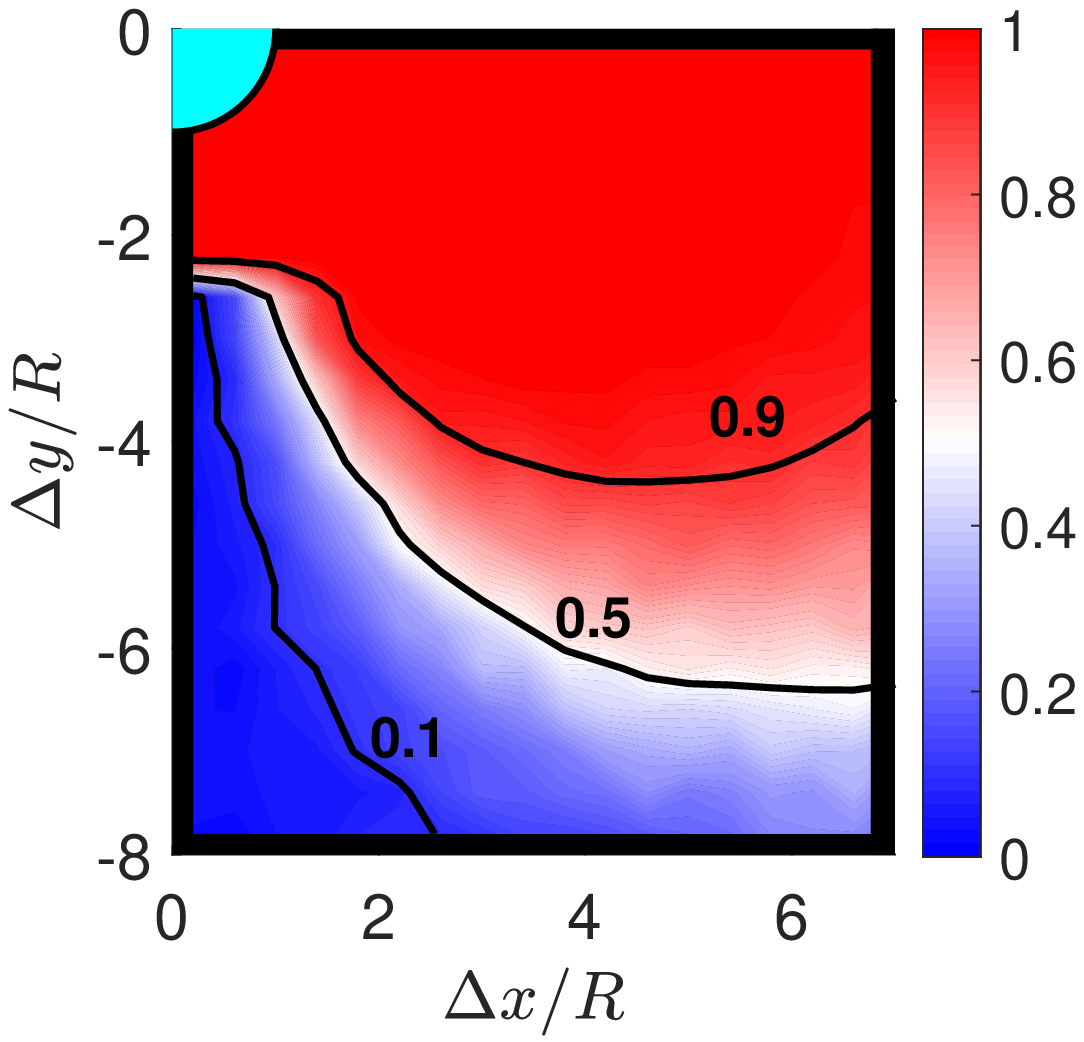}}
  \subfloat[]{\includegraphics[width=45mm,trim=35 0 65 0, clip]{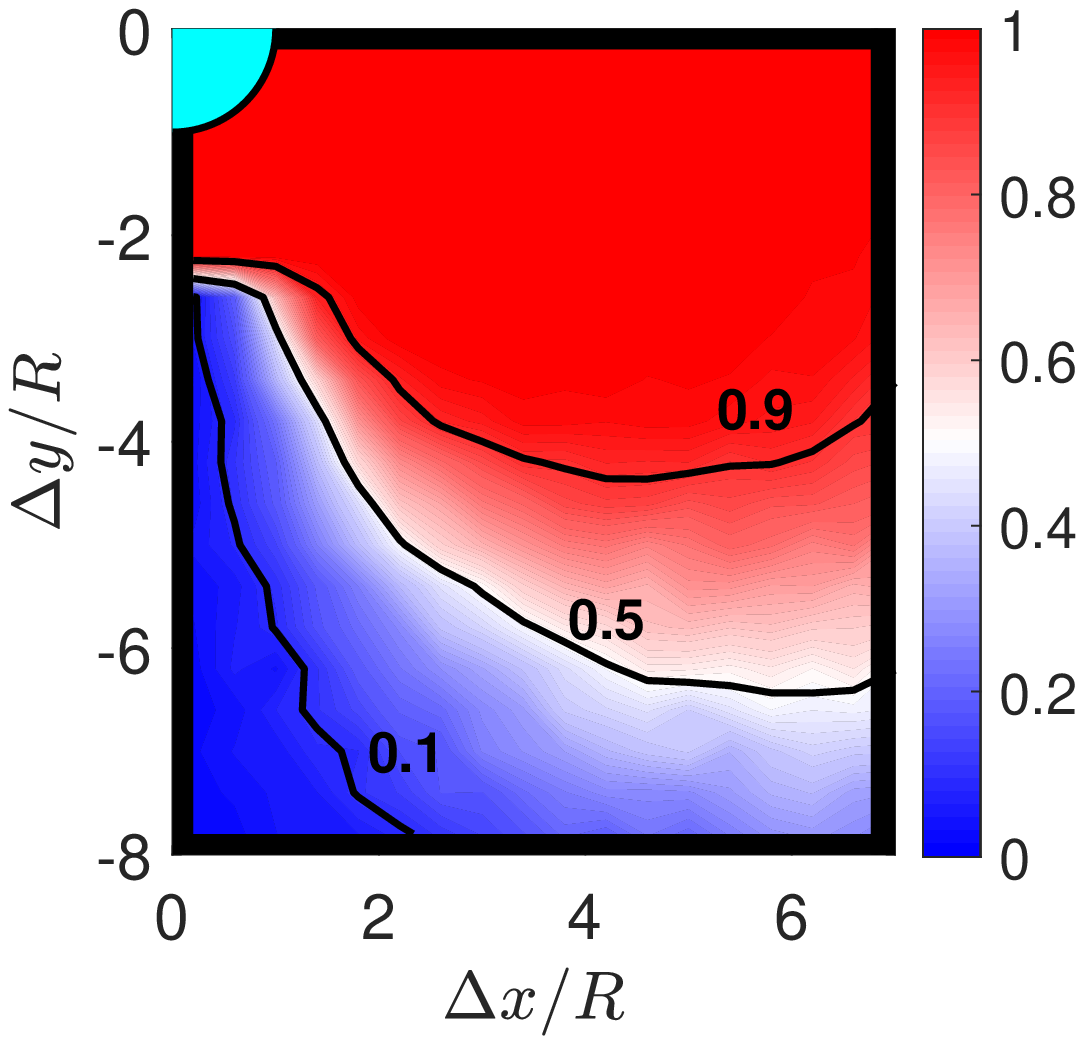}}\\
  \subfloat[]{\includegraphics[width=45mm,trim=35 0 65 0, clip]{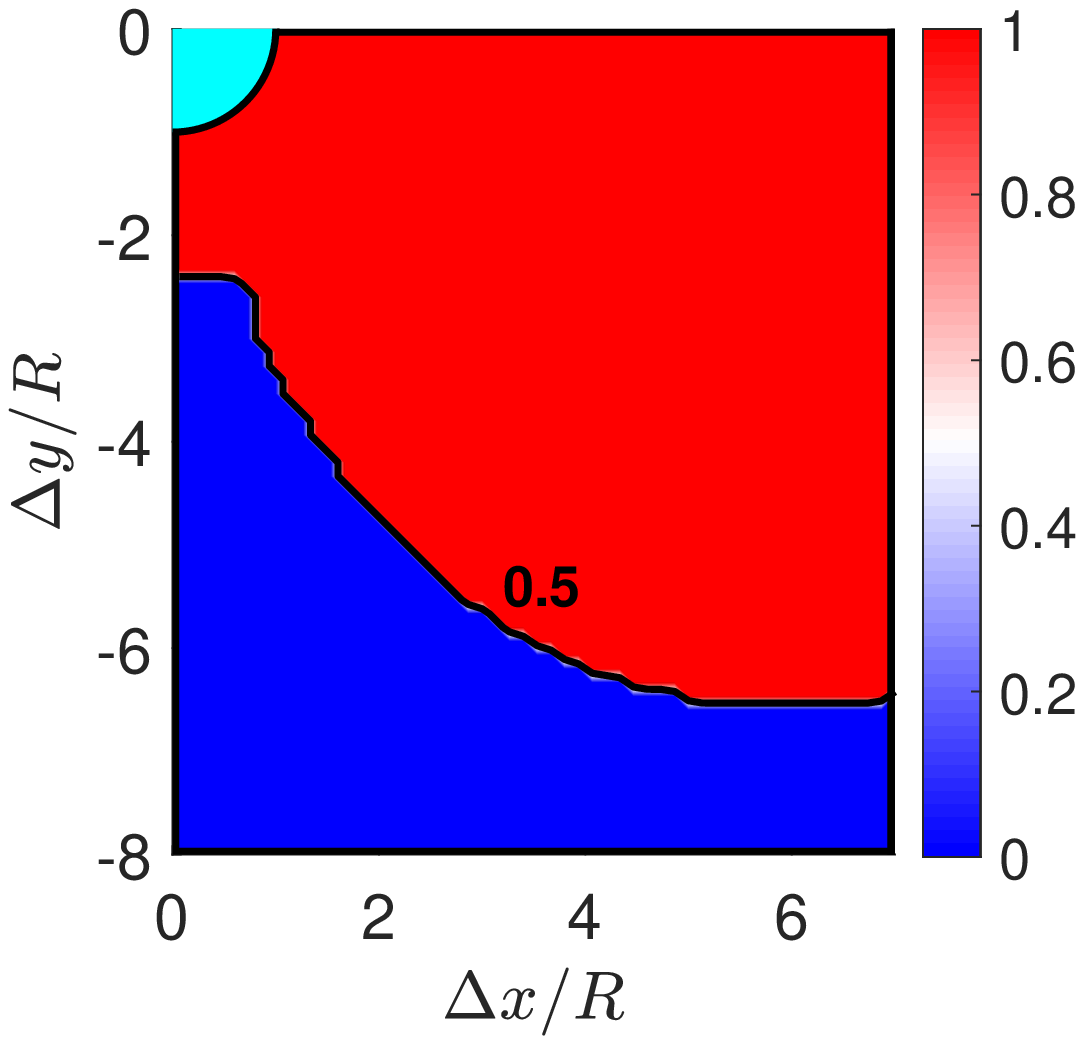}}
  \subfloat[]{\includegraphics[width=45mm,trim=35 0 65 0, clip]{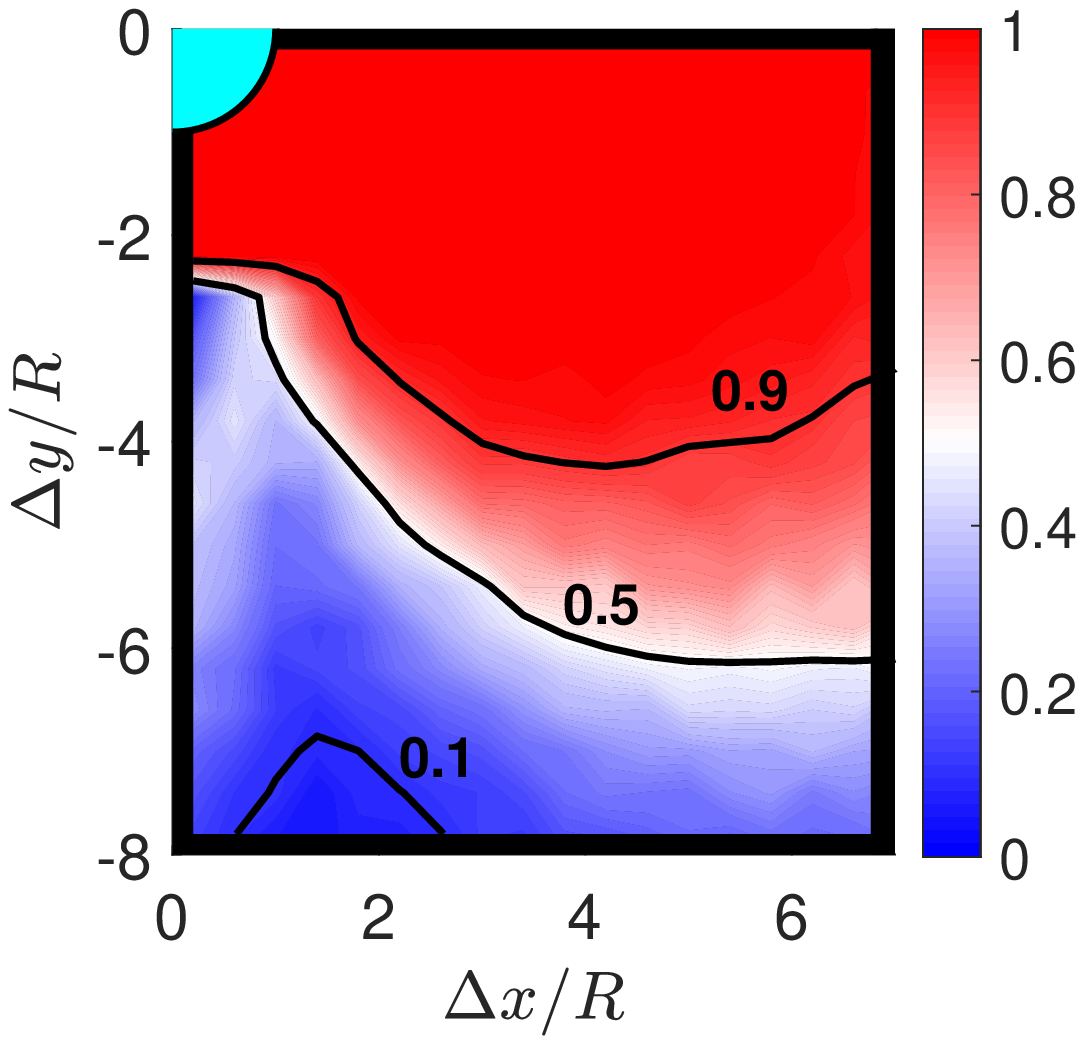}}
  \subfloat[]{\includegraphics[width=45mm,trim=35 0 65 0, clip]{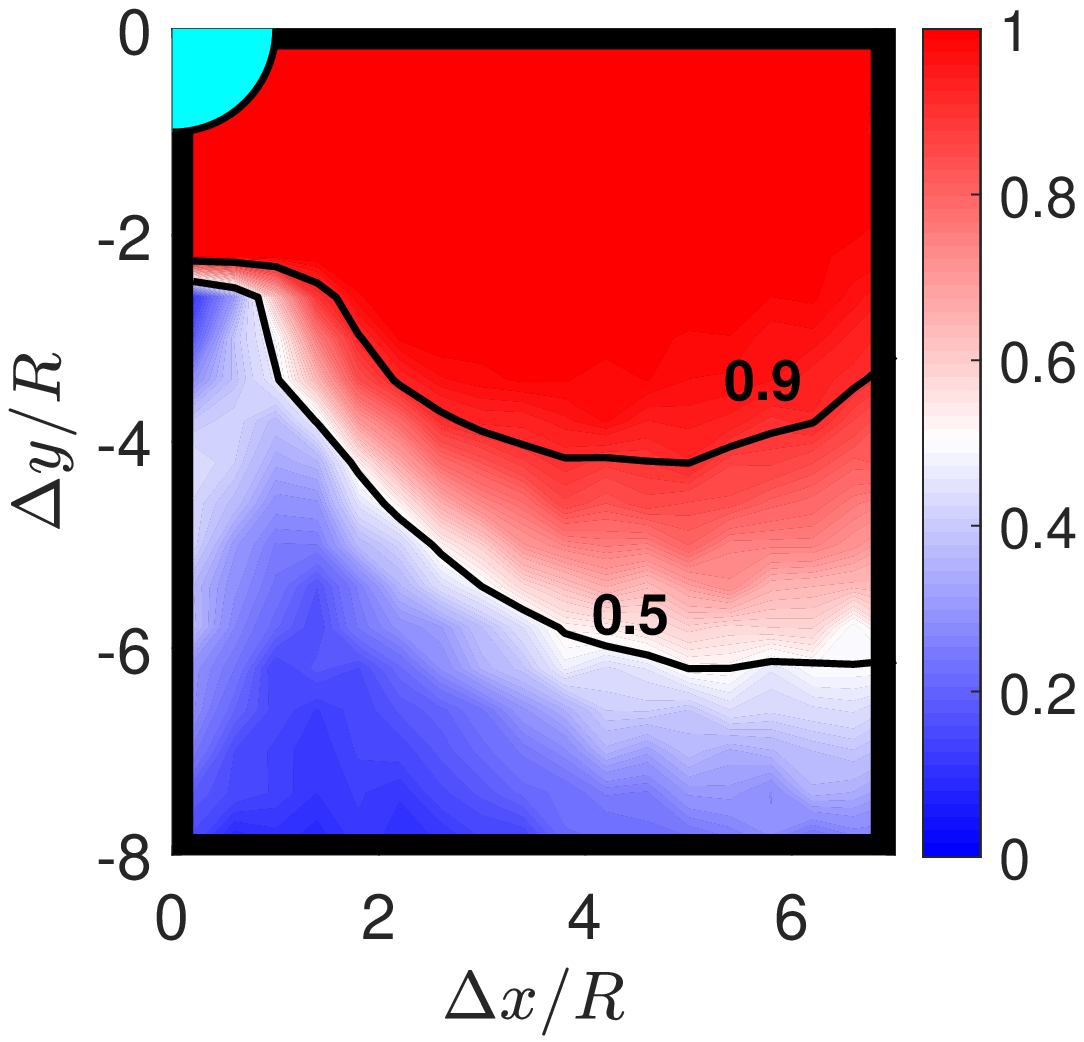}}
  \caption{Contours of the probability for pairs of bubbles to reach the side-by-side position within 1.5 s, in the case that the trailing bubble is initially placed at each given coordinate relative to the leading bubble in the domain. (a), (b), and (c) are obtained without agitaiton, with spatially uncorrelated agitation, and with spatially correlated agitation, respectively, without viscous interaction. (d-f) are contour parts of (a-c) with viscous interaction.}
\label{fig:cont}
\end{figure}
For each case in table \ref{tab:pm}, figure \ref{fig:cont}a-f show contours of the probability at $t=1.5$ s, $p_{T}(t=1.5 s,\Delta x,\Delta y)$.
Figure \ref{fig:cont}a and d show almost no difference, suggesting the small influence of the wake on the probability, in the case without agitation. Due to the absence of the agitation, the probability is binary (0 or 1) changing across the contour line.
In the plots of the cases with agitation, the contours change with a finite gradient due to the smooth change of the probability.
The contour lines of 0.5 and 0.9 are nearly common between these plots.
The lines of 0.5 in both plots are also similar to those in figure \ref{fig:cont}a and b.
Interestingly, the contour lines of 0.1 present differences among the contours.
In figure \ref{fig:cont}b and c, the contour follows the wake profile, indicating that the probability of initially in-line pairs to reach side-by-side within 1.5 s is sufficiently small.
On the other hand, in figure \ref{fig:cont}b, the contour line of 0.1 is pushed down below $\Delta y =-7R$.
In figure \ref{fig:cont}c the corresponding contour is not even present.
These changes indicate that the probability for in-line bubbles is significantly increased by the presence of the wake.
The difference between figure \ref{fig:cont}b and c, and that between figure \ref{fig:cont}e and f are small.
This result indicates that the spatial correlation of the agitation has a small influence on the probability.

\begin{figure}
\centering
  \includegraphics[width=65mm,trim=0 0 0 0, clip]{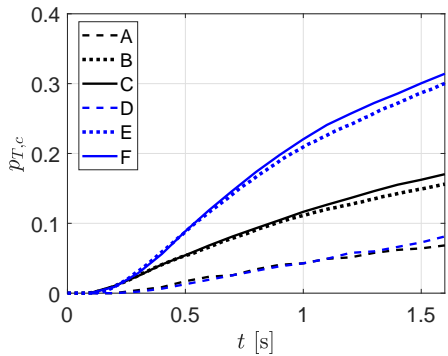}
  \caption{Evolution of the probability of side-by-side clustering in the case that the trailing bubble is initially released in the wake region, obtained from the sets shown in table \ref{tab:pm}.}
\label{fig:Pcum}
\end{figure}
Figure \ref{fig:Pcum} shows the evolution of the conditional probability of side-by-side clustering to occur for in-line configurations of pairs.
The conditional probability is defined as the probability field integrated over the wake region at each instant:
\begin{equation}
    p_{T,c}(t)=\frac{1}{\Omega_w}\int_{\Omega_w}p_c(t,\Delta x,\Delta y) d(\Delta x)d(\Delta y),
\end{equation}
where $\Omega_w$ is the wake region defined as $\Omega_w:\Delta x\in[0,2R], \Delta y\in[-8R,-2.25R]$.
For all cases, the probability monotonically increases with time, and the slopes of the plots grow up to $t\approx0.5$ s and then decay. The increase is because pairs with larger inter-bubble distance reach side-by-side at a later time.
Without agitation, regardless of viscous interaction, the probabilities reach approximately 0.075 at $t=1.5$ s (case A and D).
With agitation, viscous interaction makes difference in the probability evolution.
Without viscous interaction and spatially uncorrelated agitation (case B), the probability reaches approximately 0.15 at $t=1.5$ s. With the use of spatially correlated agitation (case C), the probability becomes slightly larger at all $t$.
With viscous interaction and spatially uncorrelated agitation (case E), the probability reaches approximately 0.3 at $t=1.5$. The spatial correlation of the agitation likewise increases the probability with a slight magnitude (case F).
Overall, compared to the cases without agitation, the probability is increased by a factor of 2 by modeling the agitation, and by a factor of 4 by modeling both the agitation and viscous interaction, regardless of the spatial correlation of the agitation, at all $t$.
These results indicate that the combination of viscous interaction and turbulence can significantly enhance the side-by-side clustering of initially in-line pairs, at the timescale of the dilute cluster formation in the experiment.

\subsection{Implication to the formation of larger clusters}
Viscous wake and potential interaction are common for spherical bubbles rising at $Re>1$, and the symmetry breaking of the in-line configurations as well as side-by-side clustering of bubbles that were identified in the previous sections could be observed in various regimes of bubbly flows.
The potential-induced attraction that leads to the formation of clusters perpendicular to gravity was predicted by previous models of three-dimensional bubbly dispersion in potential flows (\citet{Sangani93, Smereka93, Yurkovetsky96, Spelt98}).
Such strong clustering has not been observed in experiments. The discrepancy between the theory and experiment has been associated with fluctuations induced by bubbles that are not included in these theories, including path instability due to deformations of bubbles and the wake-induced disturbances in the flow \citep{Risso18}.
On the contrary, the present results suggest that, for bubbles trapped near the wall, the viscous wake may accelerate the clustering of bubbles.
\citet{Bouche12} pointed out that the viscous wake is largely attenuated when bubbles rise in a thin-gap. In the present experiment, the depth of the channel is much greater than the size of bubbles and bubbles are not subjected to such constraints.
Given the agreement between the experiment and the model, we consider that the effects of the channel geometry may not drastically influence the physics of pairwise clustering identified in the previous sections, within the validity of the model.

At a bulk void fraction of $O(1)$\% or greater, denser and larger clusters can appear, compared to the chain-like clusters shown in figure \ref{fig:cluster}. These clusters rise faster than isolated bubbles, and catch up with leading bubbles to merge and grow vertically as well as breakup \citep{Takagi11}. Detailed mechanisms of inter-bubble interactions that lead to these dynamics remain uncovered. In this regime, bubble clusters not only interact with the background turbulent flow field but also can alter the flow through two-way coupling, and description using the present model based on one-way coupling may be inaccurate. Nevertheless, during the onset of clustering from a state of random dispersion, the present approach could still be of use to predict the timescale of clustering.

\section{Conclusion}
We studied the interaction dynamics of a pair of 1 mm spherical bubbles rising near the wall in a turbulent channel flow through combined experiments and modeling, in order to shed light on the mechanism of the formation of bubble clusters in the flow.
High-speed imaging identified that pairwise bubbles tend to take nearly in-line configurations when $S>5$ and side-by-side configurations otherwise, $O(1)$ s after released in the channel.
A Lagrangian model is constructed to describe the motions of these bubbles at $Re=O(100)$ by considering hydrodynamic inter-bubble interactions and turbulent agitation. We analytically formulated the interactions through potential flow by using bi-polar expansion of the velocity potential as well as expressed interactions through bubbles' viscous wake by adopting a model of \citet{Hallez11}. We represented the turbulence-induced fluctuation of the motions of bubbles through spatio-temporal, pseudo-stochastic forcing of the bubbles.
The model was validated against the experiment by comparing the MSD and the velocity PSD of isolated bubbles, and the spatial correlations of the velocities of pairwise bubbles.
Simulations using this model identified two distinct timescales of interaction dynamics, the former of which is the rapid breaking of in-line configurations due to the shear-induced lift force acting on the trailing bubble in the viscous wake of the leading bubble, and the latter of which is the slow mutual attraction driven by potential interaction that leads to side-by-side clustering of the bubbles. The timescale of the former interaction, $O(10^{-2})$ s, is smaller than that of the turbulent agitation and also independent of the inter-bubble distance, while that of the latter interaction is greater and its fluctuation by the agitation increases with the inter-bubble distance. These dynamics were found to elucidate the observed configurations.

Statistical analyses further showed that, during the timescale of the formation of chain-like dilute clusters in the experiment, the probability of the side-by-side clustering of in-line pairs simulated with the modeled agitation and viscous interaction by up to a factor of 4, compared to those simulated without them. These results indicate that viscous interaction and turbulence may play a significant role in the onset of dilute cluster formation.
The modeled spatial correlation of the agitation, which originates from the spatial structures in background turbulence, was found to enhance the pairwise clustering with a small magnitude.
The relative importance of the effect of the spatial correlation may become significant for flows with turbulent intensity higher than that considered in the present study.
In the meantime, in bubbly flows with a greater void fraction, dense and large clusters may appear and alter the macroscopic flow structures as well as the turbulent statistics near the wall. Detailed analysis of the dynamics of these clusters may require direct flow field measurements and/or high-fidelity simulation, and is a subject of future investigations.
\\

K.M. acknowledges the Postdoctoral Fellowship Program in the Center for Turbulence Research at Stanford University, and thanks Prof. Parviz Moin for his encouragement to pursue turbulence research. The authors also thank Prof. Jun Sakakibara and Prof. Toshiyuki Ogasawara for valuable discussions. Some of the computation presented here utilized the Extreme Science and Engineering Discovery Environment (XSEDE), which is supported by NSF under grant TG-CTS190009. The experiments presented here were supported in part by the Grant-in-Aid for Scientific Research (A) (No. 21246033) and (B) (No. 21360079) of the Ministry of Education, Culture, Sports, Science and Technology (MEXT).

\appendix
\section{Formulation of potential interaction}
To derive ${C}_{\rm{D_{Int}}\it{ij}\rm pot}$ and ${C}_{\rm{L_{Int}}\it{ij}\rm pot}$, the details of the equation set is presented. By the following relation derived by \citet{Jeffrey73},
\begin{equation}
(\frac{R}{r_j})^{n+1}P^l_n(\cos{\psi_j})=\frac{1}{S^{n+1}}\sum_{m=l}^{\infty}(\frac{r_i}{d})^m{n+m\choose m+l}P^l_m(\cos{\psi_i}),
\end{equation}
$\phi_j$ can be expressed as
\begin{eqnarray}
\phi_j=\sum_{n=1}^{\infty}\sum_{m=1}^{\infty}\frac{1}{S^{n+1}}(\frac{r_i}{d})^m\{{n+m\choose m}g^{(1)}_{jn}P_m(\cos{\psi_i})\hspace{6em} \nonumber
\\ \nonumber +{n+m\choose m+1}(g^{(2)}_{jn}\cos{\varphi}+g^{(3)}_{jn}\sin{\psi})P^1_m(\cos{\psi_i)}\}\\ \nonumber
=\sum_{n=1}^{\infty}\sum_{m=1}^{\infty}\frac{1}{S^{m+1}}(\frac{r_i}{d})^n\{{n+m\choose n}g^{(1)}_{jm}P_n(\cos{\psi_i})\hspace{6em}
\\ +{n+m\choose n+1}(g^{(2)}_{jm}\cos{\varphi}+g^{(3)}_{jm}\sin{\psi})P^1_n(\cos{\psi_i)}\}.
\end{eqnarray}
Substituting this relation together with (2.8) into (2.7), we may express $\phi$ with respect to $O_i$ as
\begin{eqnarray}
\phi=\sum_{n=1}^{\infty}[\{(\frac{R}{r_i})^{n+1}g^{(1)}_{in}+(\frac{r_i}{d})^n\sum_{m=1}^{\infty}\frac{1}{S^{m+1}}{n+m\choose n}g^{(1)}_{jm}\}P_n(\cos{\psi_i})\hspace{9em}\nonumber \\
+\{(\frac{R}{r_i})^{n+1}(g^{(2)}_{in}\cos{\varphi}+g^{(3)}_{in}\sin{\varphi})\hspace{15em}\nonumber
\\
+(\frac{r_i}{d})^n\sum_{m=1}^{\infty}{n+m\choose n+1}(g^{(2)}_{jm}\cos{\varphi}+g^{(3)}_{jm}\sin{\psi})\}P^1_n(\cos{\psi_i)}].\hspace{3em}\label{phi1}
\end{eqnarray}
From symmetry, $\phi$ may be also expressed with respect to $O_j$ as
\begin{eqnarray}
\phi=\sum_{n=1}^{\infty}[\{(\frac{R}{r_j})^{n+1}g^{(1)}_{jn}+(\frac{r_j}{d})^n\sum_{m=1}^{\infty}\frac{1}{S^{m+1}}{n+m\choose n}g^{(1)}_{im}\}P_n(\cos{\psi_j})\hspace{9em}\nonumber \\
+\{(\frac{R}{r_j})^{n+1}(g^{(2)}_{in}\cos{\varphi}+g^{(3)}_{in}\sin{\varphi})\hspace{15em}\nonumber
\\
+(\frac{r_j}{d})^n\sum_{m=1}^{\infty}{n+m\choose n+1}(g^{(2)}_{im}\cos{\varphi}+g^{(3)}_{im}\sin{\psi})\}P^1_n(\cos{\psi_j)}].\hspace{3em}.\label{phi2}
\end{eqnarray}
The boundary conditions on the surface of bubble-$i$ and bubble-$j$ are given as
\begin{eqnarray} \nabla\phi\cdot\vector{n}_i&=&\vector{U}_{i}\cdot\vector{n}_i\\
&=&U_{\xi i}P_1(\cos{\psi_i})+U_{\eta i}\cos{\varphi}P_1^1(\cos{\psi_i}),
\end{eqnarray}
and
\begin{eqnarray}
\nabla\phi\cdot\vector{n}_j&=&\vector{U}_j\cdot\vector{n}_j\\
&=&-U_{\xi j}P_1(\cos{\psi_i})+U_{\eta j}\cos{\varphi}P_1^1(\cos{\psi_j}).
\end{eqnarray}
$\vector{n}_i$ and $\vector{n}_i$ are the outward normal vectors on the surfaces of bubble-$i$ and bubble-$j$. Applying these boundary conditions to (\ref{phi1}) and (\ref{phi2}), we obtain $g^{(k)}_{in}$ and $g^{(k)}_{in}$ up to $O(S^{-7})$ as
\begin{eqnarray}
g^{(1)}_{i1}&=&R(-\frac{U_{\xi i}}{2}+\frac{U_{\xi j}}{2}S^{-3}-\frac{U_{\xi i}}{2}S^{-6}+\cdots), \\
g^{(1)}_{i2}&=&R(\frac{U_{\xi j}}{2}S^{-4}-U_{\xi i}S^{-7}+\cdots), \\
g^{(1)}_{j1}&=&R(\frac{U_{\xi j}}{2}-\frac{U_{\xi i}}{2}S^{-3}+\frac{U_{\xi j}}{2}S^{-6}+\cdots), \\
g^{(1)}_{j2}&=&R(-\frac{U_{\xi i}}{2}S^{-4}+U_{\xi j}S^{-7}+\cdots), \\
g^{(2)}_{i1}&=&R(\frac{U_{\eta i}}{2}+\frac{U_{\eta j}}{4}S^{-3}+\frac{U_{\eta i}}{8}S^{-6}+\cdots), \\
g^{(2)}_{i2}&=&R(\frac{U_{\eta j}}{3}S^{-4}+\frac{U_{\eta i}}{6}S^{-7}+\cdots), \\
g^{(2)}_{j1}&=&R(\frac{U_{\eta j}}{2}+\frac{U_{\eta i}}{4}S^{-3}+\frac{U_{\eta j}}{8}S^{-6}+\cdots), \\
g^{(2)}_{j2}&=&R(\frac{U_{\eta i}}{3}S^{-4}+\frac{U_{\eta j}}{6}S^{-7}+\cdots), \\
g^{(3)}_{in}&=&g^{(3)}_{jn}=0.
\end{eqnarray}
The total interaction force exerted on bubble-$i$ due to the potential interaction, $\vector{F}_{\rm{Int}\it{ij}\rm{pot}}=\vector{F}_{\rm{D_{Int}}\it{ij}\rm pot}+\vector{F}_{\rm{L_{Int}}\it{ij}\rm pot}$, is expressed as
\begin{equation}
\vector{F}_{\rm{Int}\it{ij}\rm{pot}}=R^2
\int_0^{2\pi}d\varphi
\int_0^\pi d\psi_i \sin{\psi_i}\{\rho(-\vector{U}_i\cdot\nabla\phi+\frac{|\nabla\phi|^2}{2})\}|_{r_1=R}\ {\vector{n}_i}.
\end{equation}
By simple manipulations, we can derive relations (\ref{cgijpot4}-\ref{ceijpot7}).

\section{Formulation of stochastic forcing}
The stochastic forcing for bubble $i$ along the $x$-axis is expreesed as
\begin{align}
    \vector{F}_{W_i}\cdot\vector{e}_x
    &\approx
    \sum^{Ns}_k C(\omega^s_k)\mathrm{sin}(\omega^s_k(\mathrm{sign}(i-j)d_{ij})+\psi_{k})\sum^{Nt}_l D(\omega^t_{l})\mathrm{sin}(\omega^t_{l}t+\phi^c_{k,l}+\alpha \phi^t_{k,l}t),
\end{align}
where $\omega^s_{k}$ and $\omega^t_{l}$ are the spatial and the temporal angular frequencies of the $k$-th and $l$-th basis functions. The basis functions here are sinusoidal. $\psi_k$, $\psi^c_{k,l}$, and $\psi^c_{k,l}$ are randomly chosen within a range of $\psi_k\in[0,2\pi]$ for each $(k,l)$.
$\alpha$ controls the sharpness of the decay of the temporal spectral density.
$C$ and $D$ specify the temporal and spatial spectral densities in terms of $\omega^s_{k}$ and $\omega^t_{l}$. $C$, $D$, and $\alpha$ are chosen such that the temporal and spatial statistics of the modeled bubbles recover experimental ones. The forcing along the $y$-axis follows the same expression with distinct random variables.


\bibliographystyle{jfm}

\end{document}